\documentclass[preprint,12pt]{elsarticle}

\usepackage{a4wide}
\journal{some journal}

\usepackage{color}
\usepackage{times}
\usepackage{subfig}
\usepackage{dsfont}
\usepackage{bm, color,bbm}
\usepackage{graphicx}
\usepackage{graphics}
\DeclareGraphicsExtensions{.pdf,.png,.jpg}
\usepackage{amsbsy}
\usepackage{amsmath}
\usepackage{amsfonts}
\usepackage{amsthm}
\usepackage{float}
\usepackage[normalem]{ulem}
\usepackage{amssymb}
\usepackage{textcomp}
\usepackage{multirow}
\usepackage[subnum]{cases}
\graphicspath{{fig/}}

\definecolor{purple(html/css)}{rgb}{0.5, 0.0, 0.5}
\newcommand{\ket}[1]{| #1 \rangle}
\newcommand{\bra}[1]{\langle #1 |}
\newcommand{\bq}{\begin{quote}}
\newcommand{\eq}{\end{quote}}

\newcommand{\mi}{\mathrm{i}}

\begin{document}
\begin{frontmatter}
\title{Bloch sphere analog of qudits using Heisenberg-Weyl Operators}


\author[b1,b2]{Gautam Sharma}
%\email{gautam.oct@gmail.com}
\affiliation[b1]{organization={Center for Theoretical Physics, Polish Academy of Sciences},%Department and Organization
            addressline={Aleja Lotnikow 32/46}, 
            city={Warsaw},
            postcode={02-668}, 
            %state={},
            country={Poland}}
\affiliation[b2]{organization={Optics and Quantum Information Group, The Institute of Mathematical Sciences, a CI of Homi Bhabha National Institute},%Department and Organization
            addressline={CIT Campus, Taramani}, 
            city={Chennai},
            postcode={600113}, 
            %state={},
            country={India}}
\author[b2]{Sibasish Ghosh}
%\affiliation[first]{Center for Theoretical Physics, Polish Academy of Sciences, Aleja Lotnik\'{o}w 32/46, 02-668 Warsaw, Poland}
%\affiliation{Optics and Quantum Information Group, Institute of Mathematical Sciences, CIT Campus, Taramani, Chennai 600113, India}
%\affiliation{Homi Bhabha National Institute, Training School Complex, Anushaktinagar, Mumbai 400094, India}
%\affiliation{QpiAI India Pvt Ltd, Hub-1 SEZ Tower, Karle Town Centre, Nagavara, Bangalore - 560045, India}
\author[b1,c]{Sk Sazim}
%\email{sk.sazimsq49@gmail.com}
\affiliation[c]{organization={Institute of Physics, Slovak Academy of Sciences},%Department and Organization
            addressline={Dubravska cesta 9}, 
            city={Bratislava},
            postcode={845 11}, 
            %state={},
            country={Slovakia}}
%\affiliation{Center for Theoretical Physics, Polish Academy of Sciences, Aleja Lotnik\'{o}w 32/46, 02-668 Warsaw, Poland}
%\affiliation{RCQI, Institute of Physics, Slovak Academy of Sciences, 845 11 Bratislava, Slovakia}

%\affiliation{Optics and Quantum Information Group, Institute of Mathematical Sciences, CIT Campus, Taramani, Chennai 600113, India}
%\affiliation{Homi Bhabha National Institute, Training School Complex, Anushaktinagar, Mumbai 400094, India}
	%\email{sibasish@imsc.res.in}
	
\begin{abstract}
%Geometrical representation of physical problems is a formidable tool that we use in many instances, for example, 

%The three-dimensional Bloch sphere representation of two-level systems (qubits) is a powerful tool to visualize simple concepts, like, basic quantum operations or decoherence. However, already for a three-level system (qutrit), the state space is, in general, eight-dimensional, and has a complex structure. This eight-dimensional state space is a consequence of a particular representation of qutrits using the Gell-Mann operator basis and its close allies.  
%However, for three-level systems (qutrit), it is really difficult to understand and   
%In the representation of qudits with dimensions greater than two,
%In this work, we use the Heisenberg-Weyl (HW) operator basis to represent a qutrit that induces a four-dimensional state space. 
%This representation, however, is avoided by the community due to 
%the Heisenberg-Weyl (HW) operator basis due to 
%the presence of complex elements in the coordinate vectors. 
%To address this issue, we propose a parameterization method for qutrits using the HW operators. 
%In contrast, we find a way to circumvent this obstacle by using the properties of HW operators. 
We study an analogous Bloch sphere representation of higher-level quantum systems using the Heisenberg-Weyl operator basis. We introduce a parametrization method that will allow us to identify a real-valued Bloch vector for an arbitrary density operator. Before going into arbitrary $d$-level ($d\geq 3$) quantum systems (qudits), we start our analysis with three-level ones (qutrits). It is well known that we need at least eight real parameters in the Bloch vector to describe arbitrary three-level quantum systems (qutrits). However, using our method we can divide these parameters into four weight, and four angular parameters, and find that the weight parameters are inducing a unit sphere in four-dimension. And, the four angular parameters determine whether a Bloch vector is physical. Therefore, unlike its qubit counterpart, the qutrit Bloch sphere does not exhibit a solid structure. Importantly, this construction allows us to define different properties of qutrits in terms of Bloch vector components. 
%, we observe that it serves as a natural extension of the qubit Bloch sphere. However, 
%Unlike its qubit counterpart, the sphere in $\mathbb{R}^4$ does not exhibit a solid structure. 
We also examine the two and three-dimensional sections of the sphere, which reveal a non-convex yet closed structure for physical qutrit states. 
Further, we apply our representation to derive mutually unbiased bases (MUBs), characterize unital maps for qutrits, and assess ensembles using the Hilbert-Schmidt and Bures metrics. Moreover, we extend this construction to qudits, showcasing its potential applicability beyond the qutrit scenario.
\end{abstract}



\begin{keyword}
%% keywords here, in the form: keyword \sep keyword, up to a maximum of 6 keywords
Bloch Sphere \sep Density matrix \sep Mutually Unbiased Bases \sep Unital maps



\end{keyword}

\end{frontmatter}

%\tableofcontents


%\maketitle
\section{Introduction}
%A density matrix representing the state of a finite-dimensional quantum system of dimension $d$ is a $d\times d$ matrix, which satisfies the conditions of positive semi-definiteness, hermiticity, and trace one. In general, it is difficult to study the properties of a density matrix directly. Parametrizations of the density matrix provide a simple method to study the properties of the density matrix and to use the density matrix in solving various problems of physics. 

%One of the most famous representations of density matrices is by using the Bloch vector parametrization. 
The Bloch vector representation of two-level systems (qubit) is extremely popular because of its simplicity and its various applicability, see refs \cite{Bertlmann_2008, RevModPhys.55.855,Petruccione2012}. A qubit can be uniquely represented by a three-dimensional vector so that every point inside the Bloch sphere corresponds to a physical qubit state. This lends a simple method to not only represent the qubit states but also to identify the dynamics of the qubit. For example, all rotations of the Bloch sphere correspond to a unitary operation. However, such an extension of all the beautiful properties of the qubit Bloch sphere is not completely possible for higher dimensional states.

It is known that $d^2-1$ parameters are needed to characterize arbitrary $d$-level density matrices in $\mathbb{C}^d$ \cite{Bertlmann_2008}. Most of the works till now have used the Gell-Mann operator basis to characterize the qudits as they admit real numbers in the Bloch vector elements. This parameterization leads to $d^2-1$ dimensional geometry
which is extremely complex, and intractable even in the case of three-level systems \cite{Bertlmann_2008,KIMURA2003339,kimura2004blochvector,2006quant.ph..2065K,Menda__2006,Goyal_2016}. 
%Moreover, all the points inside the $d^2-1$ dimensional sphere do not represent physical quantum states. 
A shortcoming of this feature is that all the rotations in $\mathbb{R}^{d^2-1}$ do not represent a unitary operation, which is a prominent feature in the qubit Bloch sphere. Moreover, it is very hard to understand the general evolution of qudit using this geometry, for example, how to understand the action of unital channels in $\mathbb{C}^d$ whenever $d\geq 3$. To resolve this issue and have a qubit-like Bloch representation for higher dimensional quantum states, there have been several efforts, e.g. constructing a three-dimensional Bloch sphere representation for qutrits \cite{Eltschka_2021} and developing a multiqubit-based parametrization for qudits  \cite{10.5555/2011406.2011407,PhysRevA.93.062126}. However, these methods have their pros and cons. For instance, in the multi-qubit-based parametrization, although we get \#$\binom{d}{2}$ solid Bloch spheres  for parametrizing the quantum state space, however, the requirement to have many qubit Bloch spheres makes it difficult to study the properties of the qudit state space. Whereas, Ref. \cite{Eltschka_2021} tries to capture most of the geometric and algebraic properties of the qutrit state space via a three-dimensional representation, and it is useful in various tasks like representing the mixture of qutrit states, the unitary transformation, and the transformation under action of quantum channels. If extended to higher dimensional qudit states, this approach could be extremely useful, however, it is unclear how to extend it beyond three-level systems.  
%Moreover, the structure corresponding to physical states using the Gell-Mann operator-based representation is asymmetric with respect to different axes. Also, in the Gell-Mann operator-based representation the relation between Bloch vectors corresponding to orthonormal kets is not very useful geometrically. That is, it is not very useful for constructing an orthonormal basis starting from a pure qutrit state, by using the Bloch sphere in $\mathbb{R}^8$. 
Therefore, the features which are very prominent and useful in the qubit Bloch sphere are not present for qudits, with the currently known parametrizations using Gell-Mann operator basis.

On the contrary, the Heisenberg-Weyl (HW) operators have received much less attention because they are not hermitian and thereby require complex numbers in Bloch vector components \cite{Vourdas_2004,PhysRevA.94.010301}. As such it becomes difficult to study the parameters and put them to use. There was an attempt to address the issue of complex entries in Bloch vectors in Ref. \cite{PhysRevA.94.010301}, however, their approach uses a Hermitian operator basis constructed using HW operators such that it induces a geometry in $\mathbb{R}^{d^2-1}$. We will be comparing this with our current approach in the main text. As the HW operators do provide an alternative way to represent a quantum state, it is worthwhile to study them despite the presence of complex coefficients as there can be certain tasks where the HW operator-based representation could be more suitable, such as finding Mutually unbiased bases \cite{Bandyopadhyay2002}, understanding the properties of stabilizer states and operations \cite{1997PhDT.......232G} etc. In fact, the HW operator-based parametrization has also been used for -- a) tomography of higher-dimensional quantum states \cite{PhysRevA.94.010301}, and b) developing separability criteria for multi-qudit states \cite{Bertlmann_2008, PhysRevA.94.010301, PhysRevA.66.032319, chang2018separability}.

In this work, we use the HW operator basis to represent a qutrit, and importantly, find a way to remove the presence of complex elements in Bloch vectors. 
%try to address the issue of complex Bloch vector components while using the Weyl operator basis to parametrize a qutrit (quantum systems of dimension, $d=3$). 
In what follows, we identify four weight and four angular parameters; and observe that four weight parameters induce a unit sphere in $\mathbb{R}^4$. We also obtain the constraints on the weight and angular parameters, which give a physical qutrit density matrix. It is found that not all the points inside the sphere in $\mathbb{R}^4$ correspond to a positive semidefinite matrix. 
%But still, the new Bloch sphere has several features that look like an extension of the qubit Bloch sphere. 
To unveil the geometric structure of qutrit state space, we study its two-dimensional and three-dimensional sections completely. Our study shows that these sections are unlike those studied in previous literature (cf. \cite{Goyal_2016}). This four-dimensional geometric representation enables us to retrieve the following properties of qutrits: 
\begin{itemize}
    \item The length of the Bloch vector determines the purity of the state. It is solely determined by weight parameters.
    \item The rank of a randomly chosen qutrit state can be guessed to a certain extent. We find that the rank one states live on the surface of the unit sphere. However, the rank three states live inside the spherical ball of radius $1/2$, whereas, the rank three states live anywhere but the surface of the sphere of unit radius.   
    \item The conditions for two orthogonal or mutually unbiased vectors are quite similar to the qubit Bloch sphere under some restrictions.
    \item The Hilber-Schmidt distance between qutrit states is equivalent to a factor time of the Euclidean distance in the sphere for some states.
\end{itemize}
%It is also conjectured that the orthonormal basis kets lie on the same Bloch vector or on the antipodal points, similar to what we observe in the qubit Bloch sphere. 
Further, as a potential implication of our representation, we establish the following properties meaningfully. 
\begin{enumerate}
    \item We identify mutually unbiased bases (MUBs) in $\mathbb{C}^3$ from the geometry of the Bloch sphere in $\mathbb{R}^4$.
    \item We characterize the unital map acting on qutrit states.
    \item We find the representation of ensembles generated from Hilbert-Schmidt and Bures metric.
\end{enumerate}
%Further, we apply our representation to 1) identify mutually unbiased bases (MUBs) in 3 dimensions from the geometry of the Bloch sphere in $\mathbb{R}^4$, 2) characterize the unital map acting on qutrit states, and 3) find the representation of ensembles generated from Hilbert-Schmidt and Bures metric. 
We were able to extend our method to qudits and show its importance in finding MUBs. 

The paper is organized as follows. First, we review the HW operator expansion of a qudit in Sec.\ref{II}. Then, in Sec.\ref{III} we present the Bloch sphere in $\mathbb{R}^4$ and obtain the constraints on the parameters from one, two, and three-dimensional sections in Sec.\ref{IV}. In Sec.\ref{V}, we have studied a few features of the new Bloch sphere. The Sec.\ref{VI} describes the implications of our representation. After that, in Sec.\ref{VII} we describe a way to use a similar approach for qudits. We present a comparative discussion of our construction with that of Ref. \cite{PhysRevA.94.010301} in Sec.\ref{IX}. Finally, we conclude in sec.\ref{VIII} with a summary and future works possible based on our work.

\section{Expanding a Qudit in the Heisenberg Weyl operator basis (HW)}\label{II}
\bq
We declare here that all the operations $\{\pm, \times, \div \}$ on the index space are always congruence modulo $d$ on the set of integers. For example, see Eq. (\ref{XandZ}).
\eq
Heisenberg-Weyl operator basis is defined as $\{U_{00}=\mathbb{I},U_{pq}|p,q\in[0,d-1]$, where $U_{pq}=\omega^{-\frac{pq}{2}} Z^pX^q$.   
HW operators, $U_{pq}$, are unitary operators with several desirable properties which makes them useful in several applications \cite{Cotfas_2012,PhysRevA.74.032327,PhysRevLett.114.250403,PhysRevLett.70.1895,doi:10.1142/S0129055X03001710,PhysRevA.70.062101}. These operators are constructed from the generalized Pauli operators $X$ and $Z$, which are also referred to as boost and shift operators respectively. They can be defined by their action on a pure state in the computational basis as
\begin{align}\label{XandZ}
X\ket{n}=\ket{n+1 \text{ mod d}},\:\: Z\ket{n}=\omega^n\ket{n},
\end{align}
where $\omega=e^{2\mi \pi/d}$ is the $d$-th root of unity. 
Using HW basis, we can decompose a bounded density matrix operator in $\mathbb{C}^d$ \cite{Bertlmann_2008,PhysRevA.94.010301} as
\begin{align}\label{weylexp}
\rho=\frac{1}{d}\sum_{p=0,q=0}^{d-1}b_{pq}U_{pq}=\frac{1}{d}\big(\mathbb{I}+\sum_{(p\neq 0\neq q)}b_{pq}U_{pq}\big),
\end{align}
where $b_{00}=1$ and $b_{pq}={\rm Tr}{\rho U_{pq}^{\dagger}}$ form the Bloch vector components. However, the $b_{pq}$'s are complex in general because $U_{pq}$ are not hermitian. Hence, we must find $d^2-1$ complex numbers to characterize a state completely. One can see that for $\rho^\dagger=\rho$, the coefficients, $b_{pq}^*=e^{-2pq\pi\mi}b_{-p,-q}$. Also, the restriction ${\rm Tr}[\rho^2]\leq 1$ implies the length of vector $\bm b:=\{b_{pq}\}$ is $|\bm b|\leq \sqrt{d-1}$.

Now to summarise, we notice two crucial avenues to improve from the above formalism: -- a) Can we find a way to have real entries in Bloch vectors, and b) reduce the number of relevant parameters for Bloch sphere-like representation?. A solution for (a) was suggested in \cite{PhysRevA.94.010301} by introducing a hermitian generalization of the HW operators to make the Bloch vector components real, however, the relevant Bloch sphere parameters remained equal to $d^2-1$. 
%Moreover, it is not possible to define what is a ``Bloch" sphere for these complex components. 
In the next section, we suggest an alternate approach to address these issues.

%%%%%%%%%%%%%%%%%%%%%%%%%%%%%%%%%%%%%%

%Notice that the normalization $\sqrt{d-1}$ is to ensure that $0\leq {\rm Tr}[\rho^2]\leq 1$. 
%with $|\tilde{n}|\leq \frac{d-1}{d}$. 
%Clearly, the elements of vector $\bm n$ are determined by the following relation
%%\begin{align}
%    \tilde{n}_k=\frac{1}{\sqrt{d-1}}{\rm Tr}[\rho \tilde{H}_k].
%\end{align}
\section{$\mathbb{R}^{4}$ Bloch sphere representation of a qutrit}\label{III}
In this section, we propose a Bloch sphere-like geometric construction in $\mathbb{R}^4$ for qutrits. Using HW basis, an arbitrary qutrit can be expanded as
\begin{align}\label{qutritexp}
\rho=\frac{1}{3}\left(\mathbb{I}+b_{01}U_{01}+b_{10}U_{10}+b_{11}U_{11}+b_{02}U_{02}+b_{20}U_{20}+b_{12}U_{12}+b_{21}U_{21}+b_{22}U_{22}\right).
\end{align}  
Using the property, $\rho^\dagger=\rho$, we find that the coefficients $b_{pq}$ must obey the following relations
\begin{align}\label{Blochrelations}
&b_{01}=n_1e^{i\theta_1}, b_{02}=n_1e^{-i\theta_1},\:\:\:
b_{10}=n_2e^{i\theta_2}, b_{20}=n_2e^{-i\theta_2},\nonumber\\
&b_{12}=n_3e^{i\theta_3}, b_{21}=n_3e^{-i\theta_3},\:\:\:
b_{22}=n_4e^{i\theta_4}, b_{11}=n_4e^{-i\theta_4},
\end{align}
where $n_i,\theta_i\in \mathbb{R}$.  
Thus, we can rewrite the expansion of $\rho$ as 
\begin{align}\label{qutritexp1}
\rho=&\frac{1}{3}\left[\mathbb{I}+n_1(e^{i\theta_1}U_{01}+e^{-i\theta_1}U_{02})+ n_2(e^{i\theta_2}U_{10}+ e^{-i\theta_2}U_{20})\right. \nonumber\\ &\left.+
n_3(e^{i\theta_3}U_{12}+e^{-i\theta_3}U_{21})+ n_4(e^{i\theta_4}U_{22}+e^{-i\theta_4}U_{11})\right].
\end{align} 
%Then from Eq. (\ref{qutritexp1}), we find that eight real parameters that can completely characterize a qutrit. The unique property of these parameters is that they consist of four weight parameters $ n_i$ and four angular parameters $\theta_i$'s, with $-1\leq n_i\leq1$ and $0\leq \theta_i \leq \pi$. We define $\vec{n}=\{n_1,n_2,n_3,n_4\}$ as the four-D Bloch vector. The $n_i's$ are the weight elements corresponding to each commuting pair of HW operators. We will now obtain the constraints on the Bloch vector parameters $n_i$ and the angular parameters $\theta_i$.

%Note 1- \textit{An implication of using the four-D Bloch vector representation is that more than one state lies at the same point in the sphere. The states lying on the same point are distinguished only by the angular parameters $\theta_i$. These states are equivalent under the action of some unitary operators. It would be interesting to identify these unitary operators.} \\

%Note 2- \textit{One can also characterize a qutrit state by allowing $0\leq n_i\leq1$ and $0\leq \theta_i \leq 2\pi$ range of values, but then we will not obtain a sphere structure.} 

%Such a structure might also be worth studying because the weight parameters have proper  meaning when they are positively valued. However, we do not study this in the current work.

%%%%% New idea %%%%%%%%%%%%%%%%%%%%%%   
Now, from Eq. (\ref{qutritexp1}), we observe that one can define a set of matrices $\{H_i\}$, where $H_1=e^{\mi \theta_1}U_{01}+e^{-\mi \theta_1}U_{02}$, $H_2=e^{\mi \theta_2}U_{10}+e^{-\mi \theta_2}U_{20}$, $H_3=e^{\mi \theta_3}U_{12}+e^{-\mi \theta_3}U_{21}$ and $H_4=e^{\mi \theta_4}U_{22}+e^{-\mi \theta_4}U_{11}$. The matrices, $H_i$, are Hermitian, traceless, and ${\rm Tr}[H_i H_j]=6\delta_{ij}$ for all values of $\theta_i$. %Therefore, the set of matrices, $\{H_m\}$, forms a basis.
Then, a state in $d=3$ can be written in the following form
\begin{align}\label{qutritexpnew1}
\rho=\frac{1}{3}\left[\mathbb{I}+\bm n .\bm H\right],\:\:{\rm with}\:\: n_i=\frac{1}{2}{\rm Tr}[\rho H_i].
\end{align} 
where $\bm n$ is a real vector in $\mathbb{R}^4$ with $|\bm n|^2\leq 1$. Therefore, we find that the construction in Eq. (\ref{qutritexp1}) is analogous to the qubit Bloch sphere. We note here that the angle parameters $\theta_i$s are determining which states within the sphere in $\mathbb{R}^4$ are valid. An implication of using the Bloch vector representation in $\mathbb{R}^4$ is that more than one state lies at the same point in the sphere. The states lying on the same point are distinguished only by the angular parameters $\theta_i$. These states are equivalent under the action of some unitary operators. It would be interesting to identify these unitary operators. Later, we will shed some light on this fact. 
%We can speculate, these unitaries form a discrete $U(1)$ group.
%with $|\bm n|^2\leq \frac{2}{3}$. 
%Clearly, the elements of vector $\bm n$ are determined by the following relation
%\begin{align}
  %  n_m=\frac{1}{\sqrt{2}}{\rm Tr}[\rho H_m].
%\end{align}

%%%%%%%%%%%%%%%%%%%%%%%%%%%%%%%%%%%%%%%%%%%%%%%%%%%%%%%%%%%%%%%%%%%%%%%%%%%%%%%
\section{Constraints on the Bloch vector and angular parameters -- for $d=3$}\label{IV}
It is clear that $\rho$ is hermitian, which is guaranteed by the choice of expansion coefficients. Moreover, ${\rm Tr}[\rho]=1$ as the HW matrices are traceless except for $U_{00}=\mathbb{I}$. The only condition that remains to be satisfied is the positive semi-definiteness of $\rho$, i.e. $x_i \geq 0$, where $x_i$'s are the eigenvalues of $\rho$. In order to do this, we construct the characteristic polynomial Det($x\mathbb{I}-\rho$), of the density matrix $\rho$. The necessary and sufficient condition for the eigenvalues $x_i$ to be positive semi-definite is that the coefficients $a_i$'s of the characteristic polynomial are also positive semi-definite \cite{KIMURA2003339}. The characteristic polynomial has the following form 
\begin{align}\label{chareq}
	\text{Det}(x\mathbb{I}-\rho)=\prod_{i=1}^N(x-x_i)=\sum_{j=0}^{N}(-1)^ja_jx^{N-j}=0.
\end{align}
Notice that $a_0=1$ by definition. Now, we apply Newton's formulas to find the values of other coefficients $a_i$s(for details please see ref. \cite{KIMURA2003339}). Newton's formulas relate the coefficients $a_i$ and the eigenvalues $x_i$ as
\begin{align*}
	la_l=\sum_{k=1}^{l}C_{N,k}a_{l-k}, (1\leq l \leq N),
\end{align*}
where $C_{N,k}=\sum_{i=1}^{N}x^k$.
%\subsection{Qutrits}
%The matrix form of a qutrit can be written in terms of $n_i$s and $\theta_i$s as following
%
%\begin{widetext}
	%\begin{align}\label{qutritnewexp}
	%\rho=\frac{1}{3}\begin{pmatrix}1+2n_2 \cos(\theta_2) & n_1e^{i \theta_1}+n_3e^{-i\theta_3}+n_4e^{-i\theta_4}& n_1e^{-i \theta_1}+n_3e^{i(\theta_3-\frac{2\pi}{3})}+n_4e^{i(\theta_4+\frac{2\pi}{3})}\\ \\
	%n_1e^{-i \theta_1}+n_3e^{i\theta_3}+n_4e^{i\theta_4}&1-n_2\cos(\theta_2)-\sqrt{3}n_2\sin(\theta_2)&n_1e^{i \theta_1}+n_3e^{-i(\theta_3+\frac{2\pi}{3})}+n_4e^{-i(\theta_4-\frac{2\pi}{3})} \\ \\
	%n_1e^{i \theta_1}+n_3e^{-i(\theta_3-\frac{2\pi}{3})}+n_4e^{-i(\theta_4+\frac{2\pi}{3})}& n_1e^{-i \theta_1}+n_3e^{i(\theta_3+\frac{2\pi}{3})}+n_4e^{i(\theta_4-\frac{2\pi}{3})}&1-n_2\cos(\theta_2)+\sqrt{3}n_2\sin(\theta_2) \end{pmatrix}
	%\end{align}
%\end{widetext}
Using the results directly from ref. \cite{KIMURA2003339}, we get the following expressions for $a_i$'s in terms of $\rho$ in $d=3$ as
\begin{align}\label{coeff}
	a_0=1,\:\: a_1={\rm Tr}[\rho],\:\: a_2=\frac{(1-{\rm Tr}[\rho^2])}{2!},\:\:{\rm and}\:\:a_3=\frac{1-3{\rm Tr}[\rho^2]+2{\rm Tr}[\rho^3]}{3!},
\end{align}
whereby construction,  
%Ensuring that the terms in Eq.(\ref{coeff}) are non-negative results in positive semi-definite density matrix $\rho$. 
$a_1=1$, and $a_2\geq 0$ imposes the constraint $|\bm n|^2\leq1$. 
%\begin{align}\label{a2}
	%2!a_2=1-{\rm Tr}[\rho^2]=1-\frac{1}{3}\Bigg(1+2\bigg(n_1^2+n_2^2+n_3^2+n_4^2\bigg)\Bigg)\geq 0 \:\:
	%\implies n_1^2+n_2^2+n_3^2+n_4^2\leq1.
%\end{align}
This constraint simply states that the physical states must lie inside a sphere of radius one in $\mathbb{R}^4$. The only condition remaining to be satisfied now is $a_3\geq0$, which simplifies to the following form after simple algebra, 
%\begin{widetext}
	\begin{align}\label{a3}
		%&3!a_3=1-3{\rm Tr}\rho^2+2{\rm Tr}\rho^3\geq0 \nonumber \\
		%&\implies
    1-3|\bm n|^2+2\sum_{i=1}^4n_i^3\cos 3\theta_i
   %+n_2^3\cos 3\theta_2+n_3^3\cos 3\theta_3+n_4^3\cos 3\theta_4\right)
    +6\{n_1n_2n_3\cos(\theta_1-\theta_2+\theta_3-\pi/3)-n_1n_3n_4\cos(\theta_1-\theta_3-\theta_4)\nonumber\\
    +n_2n_3n_4\cos(\theta_2+\theta_3-\theta_4+\pi/3)+n_1n_2n_4\cos(\theta_1+\theta_2+\theta_4+\pi/3)\}\geq 0.
	\end{align}   
%\end{widetext}
In the above form, it is difficult to picture the set of valid states inside the sphere. We take the one, two, and three-dimensional sections passing through the center to get a better understanding of the allowed space inside the sphere in $\mathbb{R}^4$.
\subsection{Different sections of Bloch sphere}
The condition for weight parameters, $\sum_in_i^2\leq 1$, for qutrit implies that the induced Euclidean geometry is a sphere in $\mathbb{R}^4$. However, the restriction posed by Eq.(\ref{a3}) makes it hard to understand whether the space is solid or not. To understand it, we consider some special cases (sections). To warm up, we would start with the one-section itself to see along an axis, say $n_i$, how the angular parameter is restricting it.
\begin{figure}[!h]
\centering
   \includegraphics[scale=0.5]{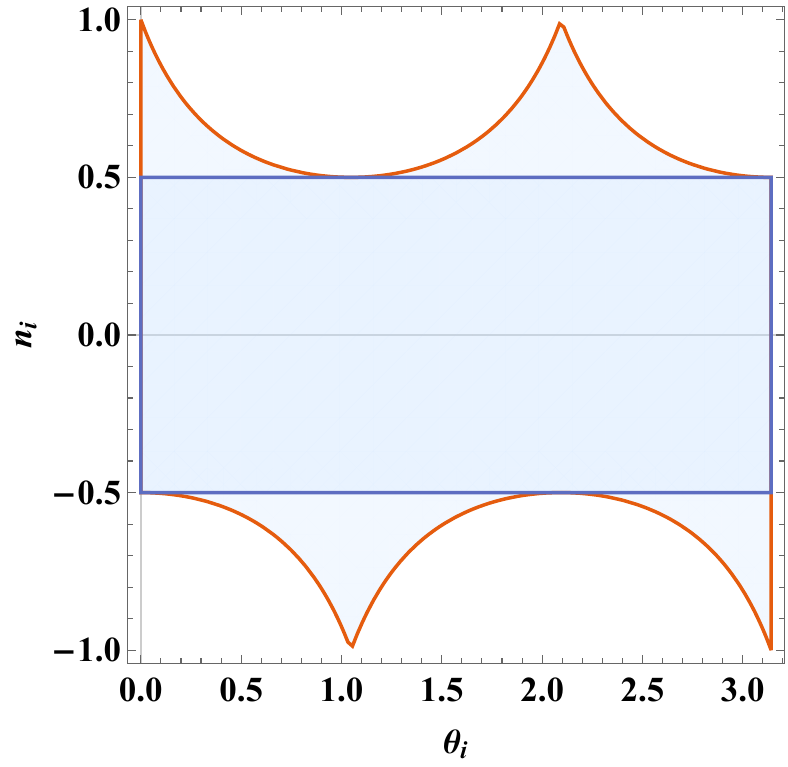}
   \caption{(Color online)The shaded region depicts the allowed values of $n_i$ and $\theta_i$ for a physical state lying on the $n_i$ axis. As can be easily seen that for $-0.5\leq n_i \leq 0.5$, all values of $\theta_i$ correspond to a physical density matrix.}
   \label{onesecpic}
\end{figure}

{\em One-dimensional sections}.--
One-dimensional sections(one section) passing through the center can be obtained by setting three out of four $n_i$'s as zero, in Eq.(\ref{a3}). We find that the expressions of one section of $a_3$ are the same with respect to all $n_i$s. Then, the condition for positivity  is given by 
\begin{align}\label{onesection}
%3!a_3^{one}=\frac{2}{9}\Bigg(
1-3n_i^2+2n_i^3\cos3\theta_i\geq 0.
\end{align}
Therefore, these one-dimensional sections are symmetric with respect to the four axes. If we rearrange the Eq.(\ref{onesection}), we get $1-n^2(3-2n\cos3\theta)\geq 0$. Clearly, $1-n^2\geq 0$ confirms that $-1\leq n\leq 1$, however, the parametric equation, $n=(3-a)/2\cos3\theta$ behaves as an envelope restricting the allowed values of $n$, where $a\in \mathbb{R}$. The one sections based on non-negativity constraint are:-
\begin{itemize}
	\item The line $-\frac{1}{2}\leq n_i\leq \frac{1}{2}$ is valid for all $ \theta_i$.
    \item The points $n_i=\pm 1$ is valid when $\theta_i$ satisfy Eq. (\ref{rank-1}). These points correspond to the pure states.
    \item The lines $\mp\frac{1}{2}\leq n_i\leq \pm 1$ are valid when $\cos3\theta_i=\pm 1$. Along these two disjoint lines, the density matrices are diagonal in the computational basis.  
%	\item The line $0.5<n_i<1$, and $-1< n_i<-0.5$ is valid for certain $\theta_i$ range as is seen in Fig. (\ref{onesecpic}).
 %$\theta \in [0,\zeta] \cup [\frac{2\pi}{3}-\zeta,\frac{2\pi}{3}+\zeta]$.
	%\item For $-1< n_i<-0.5$, $\theta \in [\frac{\pi}{3}-\zeta,\frac{\pi}{3}+\zeta] \cup [\pi-\zeta,\pi] $.
\end{itemize}
%where $\zeta=\arccos(\frac{-1}{2|n_2|})-\frac{2\pi}{3}$.  
It can be also observed (see Fig.\ref{onesecpic}) that the range of allowed values of $\theta_i$ is gradually reducing as we move away from the origin along the $n_i$, axis after $|n_i|\geq0.5$. 
\subsubsection{Two-dimensional sections}
A two-dimensional section(two sections) centered at the origin can be obtained by setting two out of four $n_i$'s to be zero in Eq.(\ref{a3}). 
%There are 6 possible two sections that can be constructed in the sphere of $\mathbb{R}^4$. 
The positivity constraint for all the two-dimensional sections have the following same form
\begin{align}\label{twosec}
%3!a_3^{two}= \frac{2}{9}\bigg(
1-3\left(n_i^2+n_j^2\right)+2\left(n_i^3\cos 3\theta_i+n_j^3\cos 3\theta_j\right)\geq0.
\end{align}
Some observation based on Eq.(\ref{twosec}) is in order. Rearranging the equation, we find $1-n_1^2(3-2n_2\cos3\theta_1)-n_2^2(3-2n_2\cos3\theta_2)\geq 0$. Clearly, in general, we have a circle of radius one, however, the parametric equations, $\{n_1=(3-a)/2\cos3\theta_1,~~ n_2=(3-b)/2\cos3\theta_2\}$ act as an elliptic envelope dictating the allowed region, where $a,b\in\mathbb{R}$. Similar to the one-dimensional sections, we see that the two-dimensional sections are also symmetric with respect to the four axes. We point out that this is unlike the Gell-Mann basis-based Bloch vector representation of a qutrit, where there exist four different types of such two sections \cite{KIMURA2003339}, which are asymmetric with respect to the axes.

\begin{figure}[!h]
	\subfloat[]{
		\includegraphics[width=0.4\textwidth, keepaspectratio]{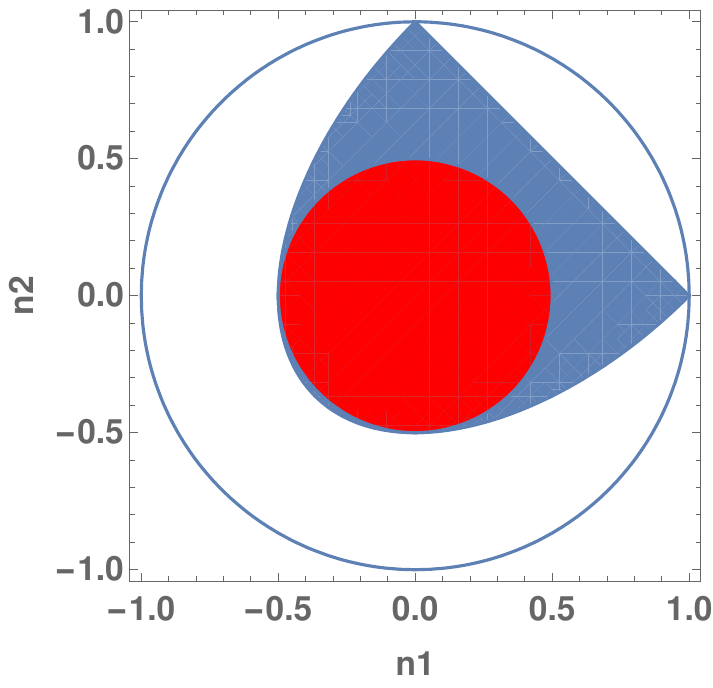}	
	}
 %\newline
 \hspace{1cm}
	\subfloat[]{
		\includegraphics[width=0.4\textwidth, keepaspectratio]{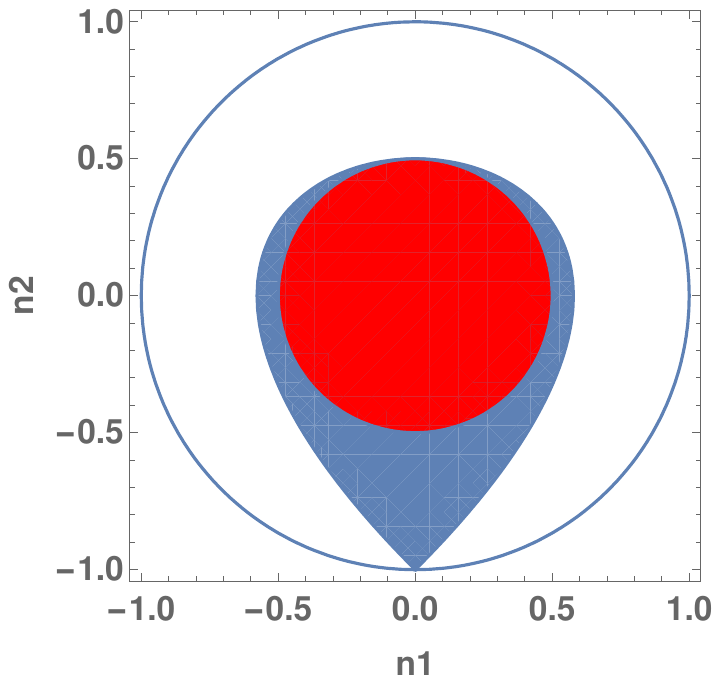}  
	}
 \newline
 \subfloat[]{
		\includegraphics[width=0.4\textwidth, keepaspectratio]{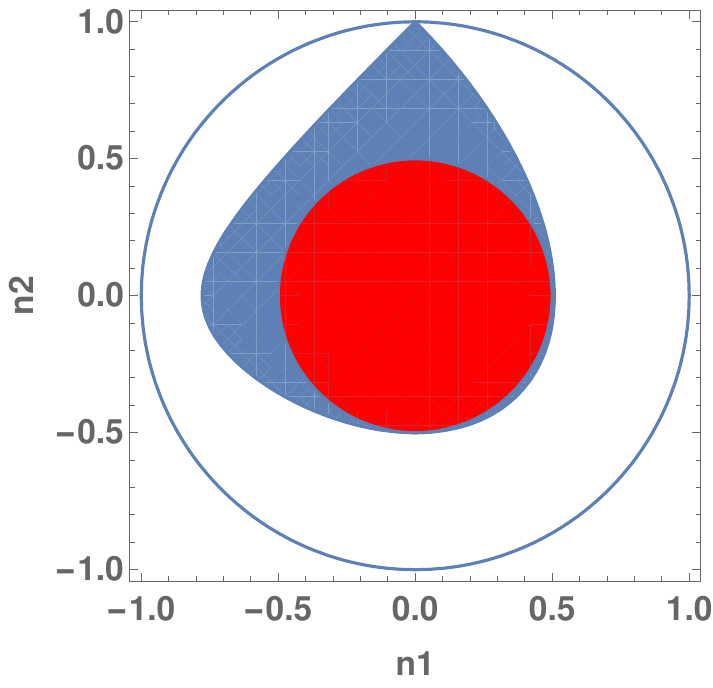}	
	}
 %\newline
 \hspace{1cm}
	\subfloat[]{
		\includegraphics[width=0.4\textwidth, keepaspectratio]{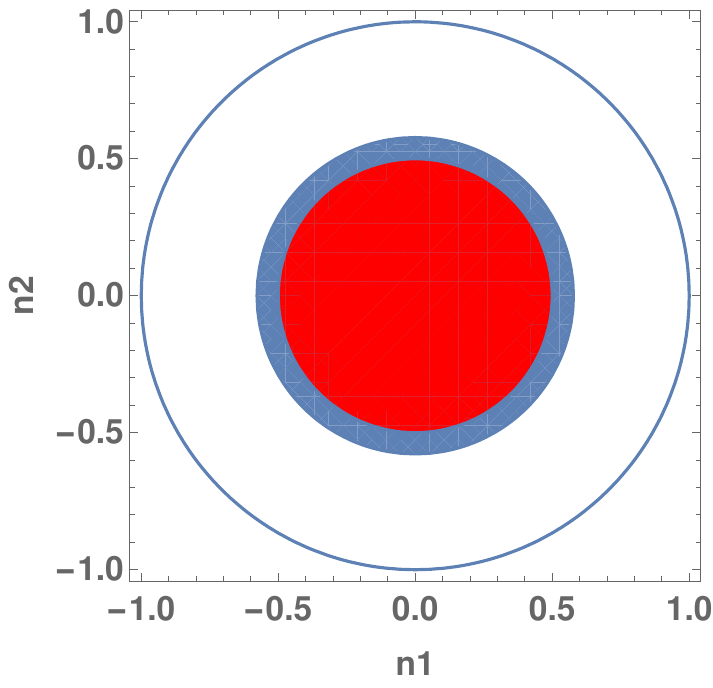}  
	}
	\caption{(Color online) Four different views of the two sections:  Blue regions are defined by -- (a) ($\cos3\theta_1=1$, $\cos3\theta_2=1$), (b) ($\cos3\theta_1=0$, $\cos3\theta_2=-1$), (c) ($\cos3\theta_1=-\sqrt{3}/2$, $\cos3\theta_2=1$) and (d) ($\cos3\theta_1=0$, $\cos3\theta_2=0$). Allowed qutrit density matrices to live inside blue regions. The red circle with a radius of $1/2$ is contained inside the blue regions in all cases.}
	\label{twosecpic}
\end{figure}

Now, we are interested in obtaining the region in the two-dimensional section which corresponds to physical qutrit states, i.e. there exist values of $\theta_i$  and $\theta_j$ so that the inequality in Eq.(\ref{twosec}) is satisfied. 
%This can be easily found by maximizing Eq.(\ref{twosec}) with respect $\theta_i$ and $\theta_j$, which gives the following inequality $1-3\left(n_i^2+n_j^2\right)+2\left(|n_i^3|+|n_j^3|\right)\geq 0$. 
Some special cases of the inequality in Eq.(\ref{twosec}) are plotted in Fig.\ref{twosecpic}, where it shows that allowed states are all inside the shaded colored regions. \\
{\em Case.1}.-- We consider ($\cos\theta_i=\pm1$, $\cos\theta_j=\pm 1$). Then, the Eq.(\ref{twosec}) reduces to 
\begin{align*}
    n_i^2(3\mp 2n_i)+n_j^2(3\mp 2n_j)=1.
\end{align*}
These four parabolas are truncated by one of the lines defined by the points $(0,\pm 1),(\pm 1,0)$ accordingly. One such parabola is shown in panel (a) of Fig.\ref{twosecpic}. \\
{\em Case.2}.-- We consider ($\cos\theta_i=0$, $\cos\theta_j=\pm 1$) or ($\cos\theta_i=\pm 1$, $\cos\theta_j=0$). Then, the Eq.(\ref{twosec}) reduces to 
\begin{align*}
    3n_i^2+n_j^2(3\mp 2n_j)=1\:\:{\rm or,}\:\:n_i^2(3\mp 2n_i)+3n_j^2=1,
\end{align*}
respectively. 
These are four ellipses stretched to one of the points $(0,\pm 1),(\pm 1,0)$ accordingly. One such ellipse is shown in panel (b) of Fig.\ref{twosecpic}. \\
{\em Case.3}.-- We consider ($\cos\theta_i=\pm \sqrt{3}/2$, $\cos\theta_j=\pm 1$) or ($\cos\theta_i=\pm 1$, $\cos\theta_j=\pm \sqrt{3}/2$). Then, the Eq.(\ref{twosec}) reduces to 
\begin{align*}
    \sqrt{3}n_i^2(\sqrt{3}\mp n_i)+n_j^2(3\mp 2n_j)=1\:\:{\rm or,}\:\:n_i^2(3\mp 2n_i)+\sqrt{3}n_j^2(\sqrt{3}\mp n_j)=1,
\end{align*}
respectively. These are four deformed parabolas akin to {\em Case.1}, touching to the circle, $n_i^2+n_j^2=1$, at only one of the points $(0,\pm 1),(\pm 1,0)$ accordingly (see panel (c) of Fig.\ref{twosecpic} for one such region). \\
{\em Case.4}.-- We consider ($\cos\theta_i=\cos\theta_j=0$). Then, the Eq.(\ref{twosec}) reduces to 
\begin{align*}
    3(n_i^2+n_j^2)=1.
\end{align*}
This is a circle of radius $1/\sqrt{3}$ which is plotted in panel (d) of Fig.\ref{twosecpic}. 

%\begin{align}\label{twosecregion}
	%{a_3}_{max}^{two}\geq0 \implies\frac{2}{9}\bigg(1-3n_i^2-3n_j^2+2|n_i^3|+2|n_j^3|\bigg)\geq 0.
%\end{align}
\begin{figure}[!h]
\centering
	\includegraphics[scale=0.8]{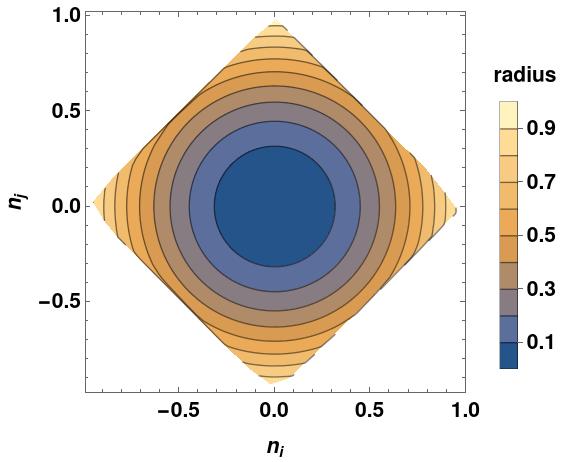}
	\caption{(Color online) {\it Two section of Qutrit state space}.-- Numerically generated qutrits satisfying Eq.(\ref{twosec}). See that within the $r=1/2$, we have concentric circles with no truncation. However, beyond $r>1/2$ concentric circles are truncated by the lines (approximated) connecting $(0,\pm 1),(\pm 1,0)$. This figure motivates us to imagine the schematic in Fig.\ref{ranks}.}
	\label{numer2sec}
\end{figure}

%\begin{figure}
	%%\includegraphics[scale=1]{fig/2sec.pdf}
	%\caption{(Color online) The region inside the blue square depicts the allowed values of $n_i$ and $n_j$ for a physical state lying on the two-dimensional section built on $n_i$ and $n_j$ axes.}
	%\label{random-2sec-state}
%\end{figure}
Further, it is informative to see the allowed values of $\theta_i$ and $\theta_j$ in different directions in the two-dimensional section as we move away from the center in the sphere ($\mathbb{R}^4$). To do this, we replace with $n_i=r\cos\alpha$ and $n_j=r\sin\alpha$ in Eq.(\ref{twosec}), so that 
\begin{align}\label{two-secr}
1-3r^2+2r^3f(\theta,\alpha)\geq0,
\end{align}
where $f(\theta,\alpha)=\cos^3\alpha\cos3\theta_i+\sin^3\alpha\cos3\theta_j$ and $-1\leq f(\theta,\alpha)\leq 1$. 
This equation captures all the allowed density matrices in the two sections. Let us list the important class of states below,
\begin{enumerate}
    \item If $r\leq 1/2$, the Eq.(\ref{two-secr}) reduces to $1+f(\theta,\alpha)\geq 0$, which is valid for all values of $\theta_i, \theta_j$ and $\alpha$. That means all the states inside this ball are valid density matrices.
    \item For $r\leq 1/\sqrt{3}$, we have $f(\theta,\alpha)\geq 0$, which means all the states inside this ball are not valid.
    \item Allowed pure states ($r=1$) implies that $f(\theta,\alpha)= 1$.
\end{enumerate}
%We note that when $r\leq0.5$, the above inequality is satisfied for all values of $\theta_i, \theta_j$ $\in$ $[0,\pi]$. 
%Moreover, to find what are the allowed values of $\theta_i$ and $\theta_j$ as we move  away from the origin in a radial direction. A 3D plot of the allowed region between $r$, $\theta_i$ and $\theta_j$ by fixing the directions is determined by $\phi$ in Fig.\ref{twosectheta}, where $\phi$ is the angle the direction makes with $n_i$. \\
To see all these items, we numerically generated $10^5$ random qutrits which satisfy Eq.(\ref{twosec}) and plotted them in Fig.\ref{numer2sec}. This again confirms our theoretical findings. 

\subsubsection{Three-dimensional sections}
Next, we consider the three-dimensional sections (three sections) centered at the origin inside the sphere (in $\mathbb{R}^4$). There are four such three-dimensional sections possible which can be obtained by setting one of the $n_i$'s as zero in Eq.(\ref{a3}). However, unlike the one and two-dimensional sections, the three-dimensional sections are all different, with the following expressions.
%\begin{widetext}
\begin{align}\label{threesec}
	&\Omega_{(n_4=0)}=1-3|\bm n|^2+2\left(n_1^3\cos3\theta_1+n_2^3\cos3\theta_2+n_3^3\cos3\theta_3\right)+6n_1n_2n_3\cos\left(\theta_1-\theta_2+\theta_3-\frac{\pi}{3}\right),\nonumber\\
	&\Omega_{(n_3=0)}=1-3|\bm n|^2+2\left(n_1^3\cos3\theta_1+n_2^3\cos3\theta_2+n_4^3\cos3\theta_4\right)+6n_1n_2n_4\cos\left(\theta_1+\theta_2+\theta_4+\frac{\pi}{3}\right),\nonumber\\
	&\Omega_{(n_2=0)}=1-3|\bm n|^2+2\left(n_1^3\cos3\theta_1+n_3^3\cos3\theta_3+n_4^3\cos3\theta_4\right)-6n_1n_3n_4\cos\left(\theta_1-\theta_3-\theta_4\right),\nonumber\\
	&\Omega_{(n_1=0)}=1-3|\bm n|^2+2\left(n_2^3\cos3\theta_2+n_3^3\cos3\theta_3+n_4^3\cos3\theta_4\right)+6n_2n_3n_4\cos\left(\theta_2+\theta_3-\theta_4+\frac{\pi}{3}\right).
\end{align} 
%\end{widetext}

\begin{figure}[!h]
	\subfloat[]{
		\includegraphics[width=0.4\textwidth, keepaspectratio]{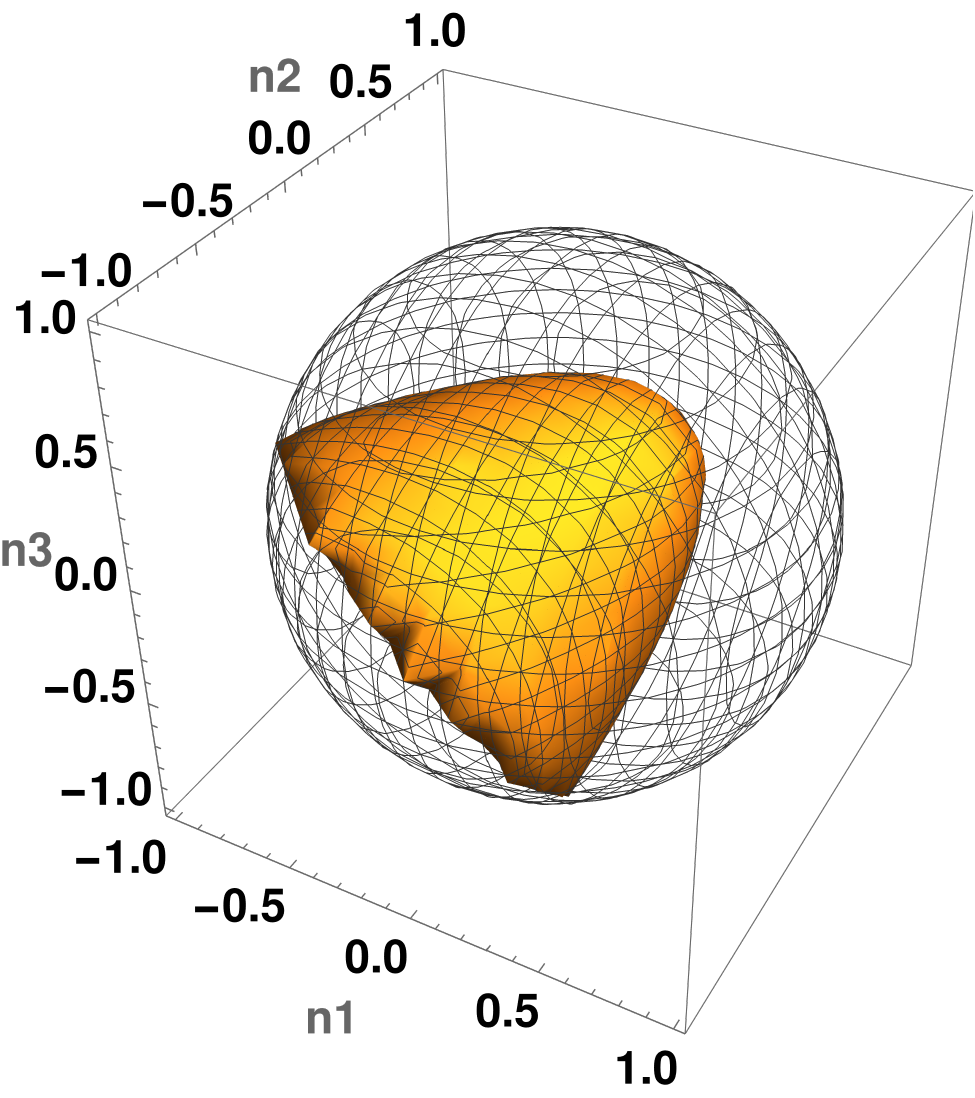}	
	}
 %\newline
 \hspace{1cm}
	\subfloat[]{
		\includegraphics[width=0.4\textwidth, keepaspectratio]{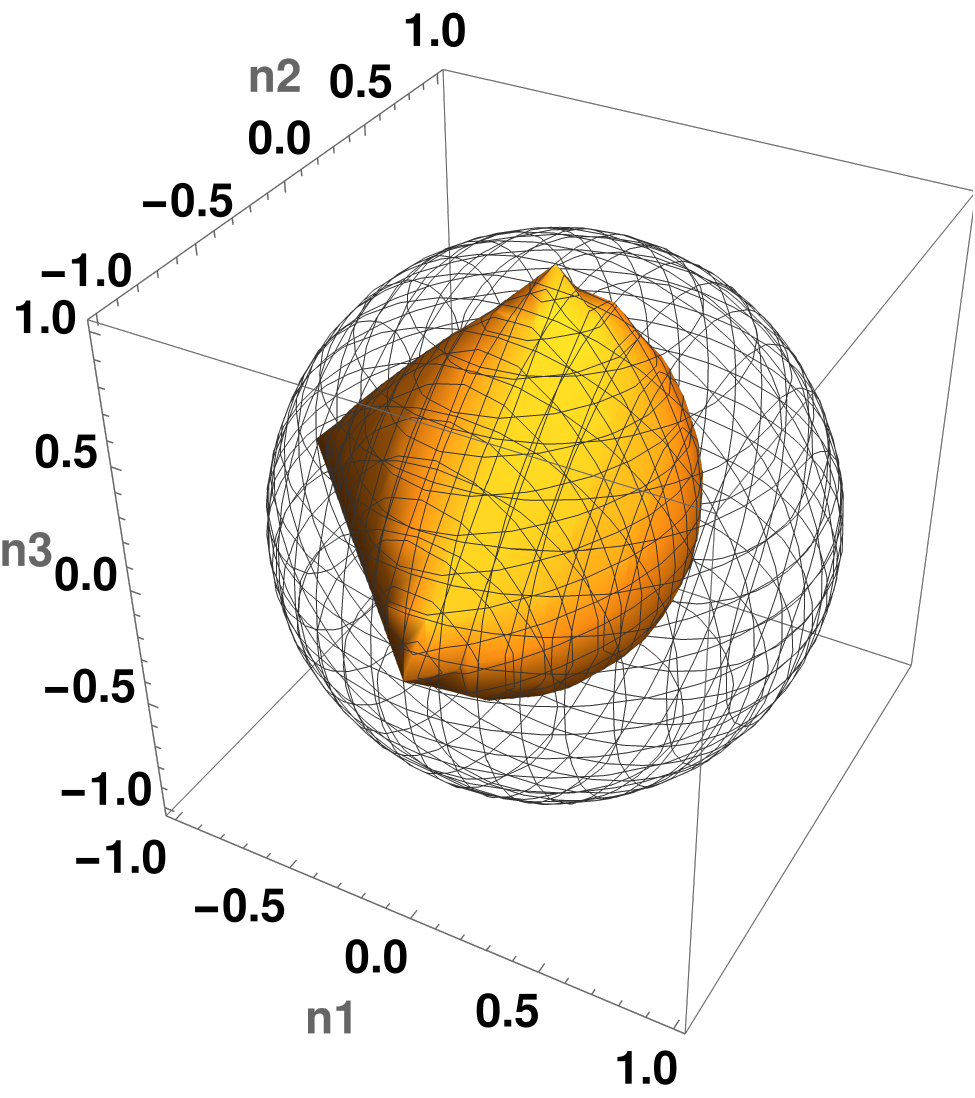}  
	}
 \newline
 \subfloat[]{
		\includegraphics[width=0.4\textwidth, keepaspectratio]{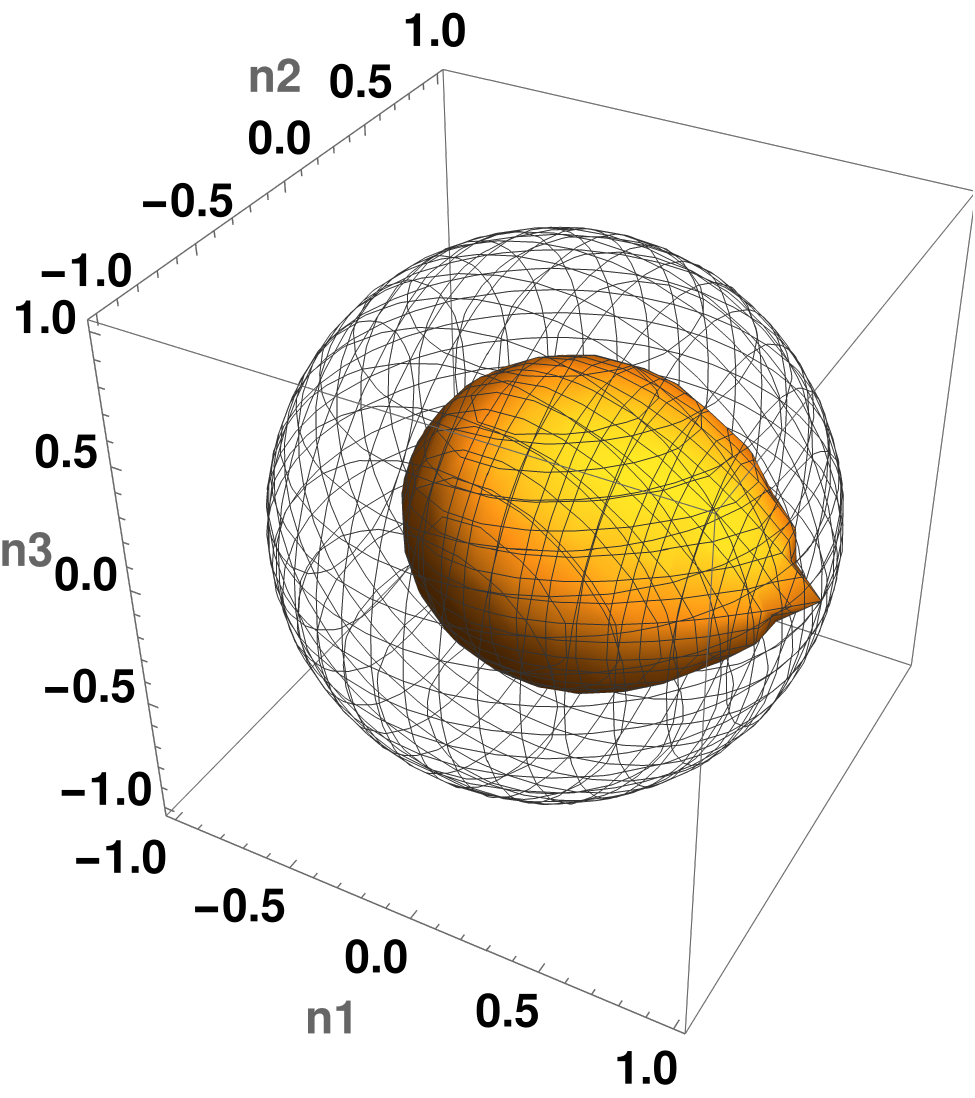}	
	}
 %\newline
 \hspace{1cm}
	\subfloat[]{
		\includegraphics[width=0.4\textwidth, keepaspectratio]{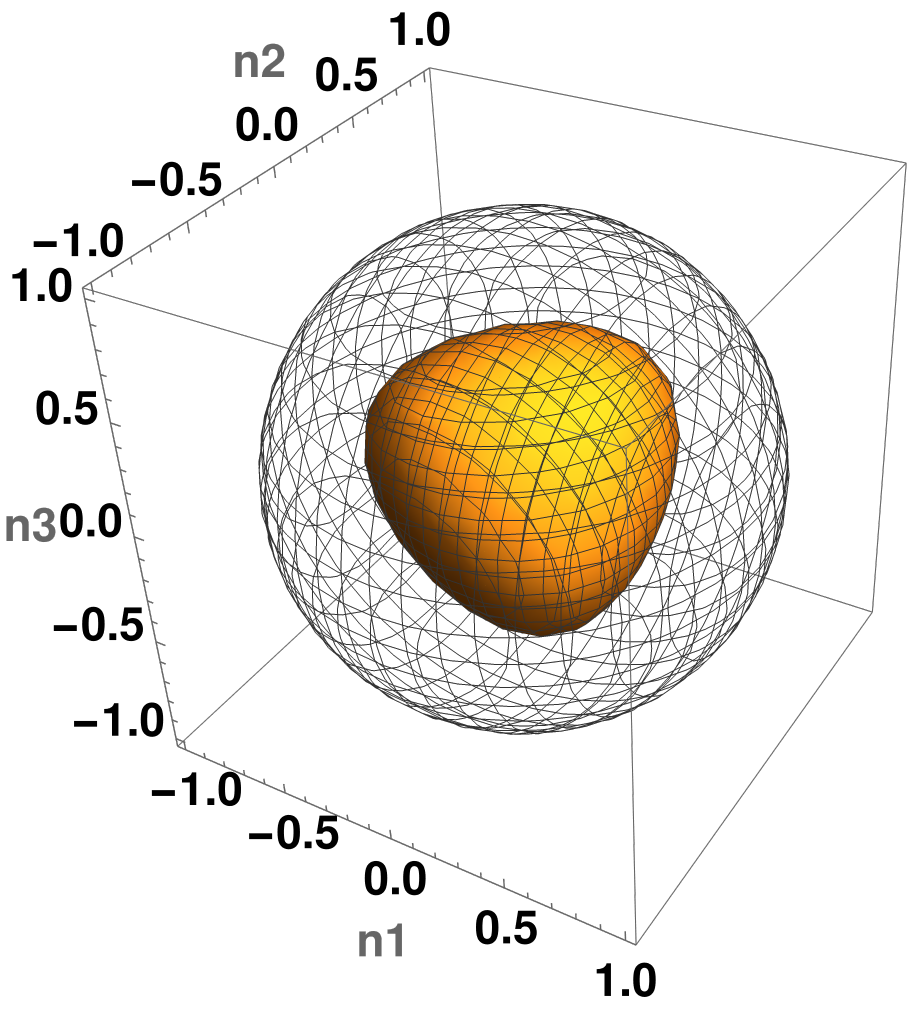}  
	}
	\caption{(Color online) Four different views of the three section $\Omega_{(n_4=0)}$: Yellow regions are defined by -- (a) ($\cos3\theta_i=-1$), (b) ($\cos3\theta_i=1$), (c) ($\cos3\theta_1=1, \cos3\theta_2=0, \cos3\theta_3=0$) and (d) ($\cos3\theta_i=0$), where $i=1,2,3$. Allowed qutrits to live inside the yellow regions. The red sphere with a radius of $1/2$ is contained inside the yellow region in all cases.}
	\label{threesecpic}
\end{figure}
It seems that these three-dimensional sections are not symmetric with respect to the axes as they have different forms in Eq.(\ref{threesec}). Therefore we need to find the regions for which the expressions in Eq.(\ref{threesec}) are non-negative. To find the non-negative regions ($\Omega_{(n_\ell=0)}\geq0$) of the three-dimensional sections means to find out whether for a given triple of $n_i,n_j$ and $n_k$, a corresponding $\theta_i,\theta_j$, and $\theta_k$ exists which gives a non-negative value of terms in Eq.(\ref{threesec}). It is difficult to do so analytically.  
%i.e., to maximize the expressions in Eq.(\ref{threesec}) with respect to $\theta$ parameters only. 
%and instead, we numerically plot the regions which satisfy $\Omega_{(n_j=0)}\geq0$. 
As we can see from Fig.\ref{threesecpic} that the different faces have different forms. WLOG, we pick three section $\Omega_{(n_4=0)}$ and plotted it for different $\theta$-values. \\
{\em Case.1}.--Panel (a) of Fig.\ref{threesecpic} depicts the three section with ($\cos3\theta_i=-1$, $i=1,2,3$). The plot reminds us of a three-dimensional parabola with bulges at different points ($n_1,n_2,n_3$). The bulges corresponding to pure states are $(-1,0,0)$, $(0,-1,0)$, $(0,0,-1)$ and $(1/3,-2/3,-2/3)$+permutations. There are other bulges corresponding to mixed states, eg., $(1/3,-2/3,-1/6)$, $(1/3,-0.744,-0.455)$, and $(1/3, 0.41,0.122)$ plus permutations. 

Notice that one will have three more similar paraboloids for the choices ($\cos3\theta_1=-1$, $\cos3\theta_2=\cos3\theta_3=1$)+permutations. The sphere of $r=1/2$ is always contained inside the paraboloid.\\
{\em Case.2}.-- If we choose ($\cos3\theta_1=\cos3\theta_2=-1$, $\cos3\theta_2=1$) instead, we get an ellipsoid with three peaks at points $(-1,0,0)$, $(0,-1,0)$ and $(0,0,1)$ which are all pure states. This ellipsoid is depicted in panel (b) of Fig.\ref{threesecpic}. The sphere of $r=1/2$ is always contained inside the ellipsoid.

Also, there are three more such ellipsoids with the choices ($\cos3\theta_1=\cos3\theta_2=1$, $\cos3\theta_3=-1$)+permutations and ($\cos3\theta_i=1$, $i=1,2,3$).\\
{\em Case.3}.-- Now the choice that two out of three $\cos3\theta_i$ is set to zero and the remaining one is equal to $\pm 1$ will yield an ellipsoid stretched to meet the sphere $n_1^2+n_2^2+n_3^2=1$ at only one point. One such example ($\cos3\theta_1=1$) is shown in panel (c) of Fig.\ref{threesecpic}. The pure state corresponding to this example is $(1,0,0)$.\\
{\em Case.4}.-- In the panel (d) of Fig.\ref{threesecpic}, the ellipsoid is considered when all ($\cos3\theta_i=0$, $i=1,2,3$). Then the ellipsoid is the generalization of circle $3(n_1^2+n_2^2)=1$ (see, two-section {\it Case.4}), as,
\begin{align*}
    3(n_1^2+n_2^2+n_3^2-\sqrt{3}n_1n_2n_3)=1.
\end{align*}
However, the circle of radius $1/\sqrt{3}$ is generalized to an ellipsoid instead of a sphere. A sphere of $r=1/2$ is contained inside this ellipsoid also. The points at which the ellipsoid is peaked are $(\pm 1/\sqrt{3},0,0)$ plus permutations.   
%we conclude that the three-dimensional sections are not symmetric with respect to the axes. 

%It can be seen from Fig.\ref{threesecpic} that the three section is of the form with bulges on the faces of an octahedron. The other three sections also are of a similar form but not the same, which we have not put in the paper for brevity. 

A few remarks from the study of the one, two, and three sections are in order. 
\begin{enumerate}
	\item It is possible to approximately construct the three sections from the knowledge of the two sections, which is not the case in the representation using Gell-Mann operator-based representation. 
 %This is because by rotating the two sections along one of the axes, one will obtain an octahedron which is a very good approximation to all three sections.
	\item It looks like from the numerical plots, that the three sections' structure is not convex. This could be because of the presence of complex coefficients. 
	\item It is also clearly visible how the one section arises from the two sections and the two sections from the three sections.
\end{enumerate}

Based on the above studies, we are ready to state the following fact of qutrit state-space in $\mathbb{R}^4$.

\textit{\textbf{Observation}}- All the points inside spherical Ball of radius $r\leq 1/2$ are physical states for all the angular parameter values of $\theta_i$'s. However, all points beyond $r>1/2$ are not valid qutrits.

Proof of this fact has been furnished in \ref{appC}. 
An implication of this result is that a rotation in the Bloch sphere does not always correspond to a unitary operation, unlike the qubit Bloch sphere. 

%This result is significant, as all the points lying inside the sphere of radius 0.5 correspond to physical states.

\subsection{Features of the Bloch sphere -- for $d=3$}\label{V}
In this section, we discuss several features of the Bloch sphere for qutrits and discuss the difference with the qubit Bloch sphere.
\subsubsection{Mixed and Pure states}
The purity of a density matrix operator is defined as  
\begin{align}
	{\rm Tr}[\rho^2]=\frac{1}{3}\Bigg(1+2\bigg(n_1^2+n_2^2+n_3^2+n_4^2\bigg)\Bigg).
\end{align}

Thus, we find that the length of the Bloch vector determines the purity of the qutrit state. Further, ${\rm Tr}[\rho^2]=1$ for $n_1^2+n_2^2+n_3^2+n_4^2=1$, i.e., the pure states lie on the surface of the unit sphere. Also,  ${\rm Tr}[\rho^2]=0$ only when $n_1^2+n_2^2+n_3^2+n_4^2=0$, i.e., the maximally mixed state lies at the center of the sphere. Also, the purity increases as we move away from the center of the sphere.

To characterize the set of pure states in $d=3$, one needs to find the states which satisfy $\rho^2=\rho$. As $\rho^2$ is a Hermitian matrix, we find its components by considering the terms $n^{(2)}_\ell={\rm Tr}[\rho^2 H_\ell]/\sqrt{2}$ and one of the elements is given by,
\begin{align}\label{eq23}
    n^{(2)}_\ell=\frac{1}{6}\left\{4n_\ell+2n^2_\ell\cos 3\theta_\ell+2f(\Bar{n}_\ell, \bm \theta)\right\},
\end{align}
where $f(\Bar{n}_2,.)=n_1n_3 \cos(\theta_1-\theta_2+\theta_3+\pi/3)+n_1n_4 \cos(\theta_1+\theta_2+\theta_3-\pi/3)+n_3n_4 \cos(\theta_2+\theta_3-\theta_4-\pi/3)$ and so on. Therefore, suitable conditions on $n_\ell$ and $\theta_\ell$ determine the set of pure states.  For example, if $\Bar{n}_\ell=0$, then the condition for pure density matrix reduces to 
%\begin{align*}
   % \rho^2=\rho \implies n_\ell^2=\pm n_\ell \:\:{\rm if}\:\: \cos3\theta_\ell=\pm 1.
%\end{align*}
%The exact solutions of the above equation are 
\begin{align}\label{rank-1}
    \rho^2=\rho \implies \begin{cases}
    n_\ell=1 \:\:{\rm and}\:\: \theta_\ell=\frac{2m\pi}{3},\\
    n_\ell=-1 \:\:{\rm and}\:\: \theta_\ell=\frac{(2m\pm 1)\pi}{3},
\end{cases}
\end{align}
where $m\in\mathbb{Z}^+$.
These states live on the boundary of sphere $|\bm n|=1$ (outer sphere). However, these states are not only extremal states. Some other solutions from Eq.(\ref{eq23}) might yield pure states (see examples in the three-dimensional section).

It is safe to assume that the states that live on the boundary of the states space are singular, i.e., ${\rm Det}[\mathbb{I}+\bm n. \bm H]=0$. It is difficult to understand the structure of the boundary from the expression of the determinant. However, it is clear that to $\mathbb{I}+\bm n. \bm H$ to be singular, $\bm n. \bm H$ should have eigenvalues equal to $-1$. Now, tracelessness of $\bm n. \bm H$ forces other two eigenvalues to $\lambda, 1-\lambda$. Also, we can easily verify that the restriction $|\bm n|\leq 1$ implies that the square of eigenvalues of $\bm n. \bm H$ is bounded by $6$, i.e., 
\begin{align*}
    1+\lambda^2+\left(1-\lambda\right)^2\leq 6,
\end{align*}
which forces $\lambda$ to be $-1\leq \lambda \leq 2$. It should be noted here that the eigenvalues of the matrices $H_\ell$ lie in the same range. Now, norm of $\bm n$ is 
\begin{align}
    |\bm n|=\sqrt{\frac{1}{6}{\rm Tr}[(\bm n. \bm H)^2]}=\left\{\frac{1}{6}\left[1+\lambda^2+\left(1-\lambda\right)^2\right]\right\}^{\frac{1}{2}},
\end{align}
whose minimum is $1/2$ when $\lambda=1/2$. This is exactly the midpoint of the values $\lambda=-1, 2$ at which $|n|=1$. Hence, if $\bm n$ is a boundary point then $-(1/2)\bm n$ is also a boundary point, heralding that the boundary points of the outer sphere ($|n|=1$) are dual to the boundary points of the inner sphere ($|n|=1/2$).

Now one can easily see that there exists another sphere for which $\bm n. \bm H$ is also  singular, i.e., for $\lambda=0$ or $1$. And midpoint of these values also defines the inner sphere ($|n|=1/2$). With these $\lambda$ values, one finds a new sphere of radius $|n|=1/\sqrt{3}$ which is self-dual, i.e., antipodal point of $\bm n$ is $-\bm n$.
\subsubsection{Rank of a Qutrit state}
\begin{figure}[!h]
\centering
	\includegraphics[scale=0.5]{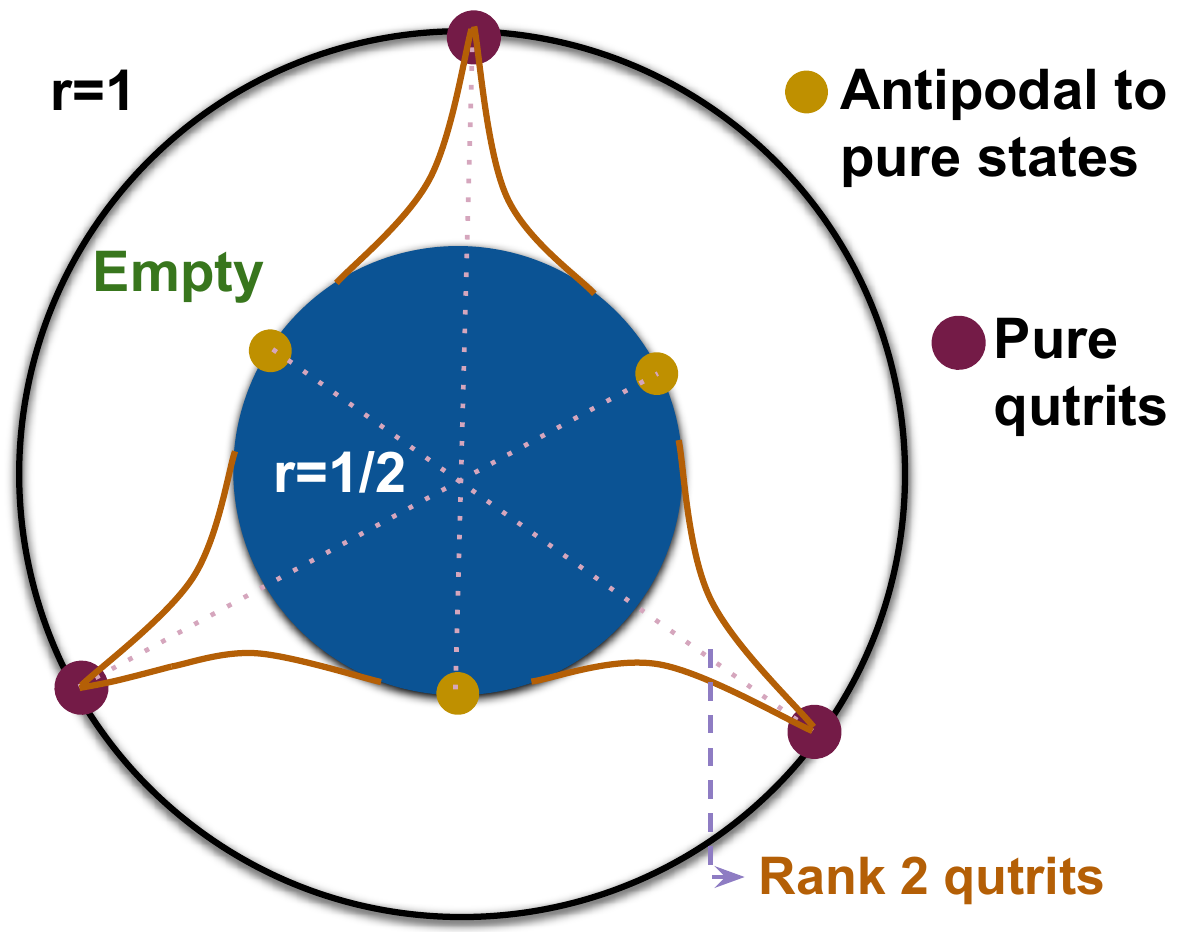}
	\caption{(Color online) {\it $2D$ projection of Qutrit state space}.-- The blue Ball ($r=1/2$) contains rank $3$ states and rank $2$ on its surface. Points inside the orange curve regions depict rank $2$ qutrits. Red points are the pure qutrits ($r=1$). The other regions are empty. (Note that it is a representative figure only for understanding qutrit state space, not the actual one.) }
	\label{ranks}
\end{figure}

A closely related concept to purity/mixedness is the rank of a physical state. Let us now recall the following equation which is equivalent to ${\rm Det}(\rho)$, 
\begin{align*}
	\Omega=1-3r^2+2r^3f(\theta_1,\theta_2,\theta_3,\theta_4,\alpha_1,\alpha_2,\alpha_3,\alpha_4).
\end{align*}
As $f(\cdot,\cdot)\in [-1,1]$, we find that $\Omega>0$ has a unique solution, i.e., $r\leq 1/2$. Notice also that the surface of the Ball ($r=1/2$) corresponds to $\Omega>0$ as well as $\Omega=0$.  Therefore, some rank $2$ qutrits also live on the surface of this Ball. Now rank $1$ and $2$ qutrits corresponds to $\Omega=0$. And we know that rank $1$ states are all situated on the surface of a sphere in $\mathbb{R}^4$ (see Eq.\ref{rank-1}). The following list summarizes our findings (see also the Fig. \ref{ranks}),
%From the qubit Bloch sphere, it is very easy to determine that the rank 1 states lie on the surface while all the remaining states are of rank 2. In our representation of qutrit states, all the pure states lie on the surface, hence they have rank 1. The challenging task is to determine where the rank 2 and 3 states lie inside the four-D sphere. To do this we note that if the determinant of a qutrit is greater than zero, i.e., $Det(\rho)>0$, then the state must be of rank 3. From the Eq.(\ref{qutritnewexp}), we find that $Det(\rho)=a_3$.
%But we saw in Lemma 3 that for $r<0.5$, $a3=Det(\rho)>0$. Hence all the states inside the four-D sphere with radius $r<0.5$ are rank 3 states. However, the states on the surface of this sphere with $r=0.5$ can be both ranked 2 and 3. Further, when $0.5\leq r <1 $ the states can be both of rank 2 and rank 3 depending on the choice of the angular parameters $\theta_i$. To summarize
\begin{itemize}
    \item Surface of sphere in $\mathbb{R}^4$ ($r=1$) contains ${\rm rank}$ $1$ qutrits.
    \item Region $\frac{1}{2}\leq r <1$ contains all ${\rm rank}$ $2$ qutrits.
      \item Inside the Ball $r\leq \frac{1}{2}$ all of ${\rm rank}$ $3$ qutrit lives.
\end{itemize}
%A representative figure of the qutrit state space depicting the location of rank 1, 2, and 3 states can be seen in Fig.(\ref{ranks}). This helps us to identify the rank of the qutrit states by simply looking at where they lie inside the four-D sphere.

\subsubsection{Orthogonal states and Mutually unbiased states}
Let us consider two pure states $\rho_1: \:(\bm n, \bm\theta)$ and $\rho_2:\:(\bm m,\bm\phi)$ and expand it in the form of Eq.(\ref{qutritexp1})
\begin{align*}
\rho_1=\frac{1}{3}\left(\mathbb{I}+\sqrt{2}\bm n. \bm H(\theta) \right),\:\: {\rm and}\:\:
\rho_2=&\frac{1}{3}\left(\mathbb{I}+\sqrt{2}\bm m. \bm H(\phi) \right).
\end{align*} 
%The states $\rho_a$ and $\rho_b$ have Bloch vectors $\vec{n_a}=\{n_{a1},n_{a2},n_{a3},n_{a4}\}$ and $\vec{n_b}=\{n_{b1},n_{b2},n_{b3},n_{b4}\}$ respectively. 
The orthogonality condition can simply be checked by ${\rm Tr}[\rho_1\rho_2]=0$. In general, the orthogonality condition for two qutrits is given by 
\begin{align}\label{orthogonalG}
\sum_{i=1}^4\cos(\theta_i-\phi_i) n_im_i=-\frac{1}{2}=\cos \left( \frac{2\pi}{3}\right). 
\end{align}
This condition is far more complex than the qubit case. 
%For example, the qutrits: $(n_1=1,\theta_1=0)$ and $(m_1=1/2,\phi_1, m_i,\phi_i|i=2,3,4)$ are othogonal whenever $\phi_1=\pm\pi+2p\pi$, where $p\in \mathbb{Z}$. 
The most simple solution exists whenever $\cos(\theta_i-\phi_i)=1$ $\forall i$, and in that case the condition \eqref{orthogonalG} reduces to 
\begin{align}\label{orthogonal}
\bm n.\bm m=-\frac{1}{2}=\cos \left( \frac{2\pi}{3}\right).
\end{align}
%This condition is richer than the orthogonality condition that we know for the two orthogonal qubit states. 
In this context, we remind our reader that Bloch vectors for two orthogonal qubit states obey, $\bm n\cdot\bm m=-1=\cos(\pi)$. 
%The orthogonality condition in Eq.(\ref{orthogonal}) reduces to $\bm n.\bm m=\cos(\frac{2\pi}{3})$.
%, whenever $\theta_{ai}=\theta_{bi}$. However, it is also needed that the points corresponding to $\vec{n_a}$ and $\vec{n_b}$ with an overlap of $\cos(\frac{2\pi}{3})$ correspond to a positive semidefinite matrix for $\theta_{ai}=\theta_{bi}$, which we were unable to identify. For $\theta_{ai}\neq\theta_{bi}$, the orthogonality condition is far more complex. 

For two mutually unbiased state vectors in $\mathbb{C}^3$, ${\rm Tr}[\rho_1\rho_2]=1/3$. Generally, two mutually unbiased qutrit will satisfy, 
\begin{align}\label{mubG}
\sum_{i=1}^4\cos(\theta_i-\phi_i) n_im_i=0=\cos \frac{\pi}{2}. 
\end{align}
Notice that if $\cos(\theta_i-\phi_i)=t$ for all $i$, where $t\neq 0$ is a real number, then it implies 
\begin{align}\label{mub}
\bm n.\bm m=0=\cos \frac{\pi}{2}.
\end{align}
Therefore, in this case, the Bloch vectors corresponding to mutually unbiased state vectors are orthogonal to each other,  which is similar to mutually unbiased qubits. 
%The Eq.(\ref{mub}) is also the condition for pure qutrits to be MUBs.  

\subsubsection{Distance between density matrices}
Let us consider two states $\rho_1:\: (\bm n, \bm \theta)$ and $\rho_2: \: (\bm m, \bm \phi$). The Hilbert-Schmidt (HS) distance between them is defined as \cite{wilde_2013}
\begin{align}
	D_{HS}^2(\rho_1,\rho_2)=\left({\rm Tr}[\rho_1-\rho_2]^2\right). %\label{HS} 
    %&=\sum_i(n_i-m_i)^2.
\end{align}
%where factor $d/(d-1)$ has been introduced to normalise the HS distance, i.e., $0 \leq D_{HS}(\rho_1,\rho_2)\leq 1$. 
HS distance defines the distance between two density matrices in induced Euclidean space \cite{PhysRevA.17.1249}. 
We obtain the Hilbert-Schmidt distance between two arbitrary qutrits,
\begin{align}\label{HS}
	D^2_{HS}[\rho_1,\rho_2] =\frac{2}{3}\sum_{i=1}^4\left\{n_i^2+m_i^2-2n_im_i\cos(\theta_i-\phi_i)\right\}.
\end{align}
This induced distance depends on the angular parameters nontrivially. 
Whenever $\cos(\theta_i-\phi_i)=1$ $\forall i$, the Hilbert-Schmidt distance reduces to the Euclidean distance in $\mathbb{R}^4$, i.e. 
\begin{align*}
    D_{HS}[\rho_1,\rho_2]=\sqrt{\frac{2}{3}\sum_i(n_i-m_i)^2}.
\end{align*}
In the qubit Bloch sphere also, the Hilbert-Schmidt distance between two density matrices is proportional to the Euclidean distance between them \cite{wilde_2013}.
%Whenever $\cos(\theta_{ai}-\theta_{bi})=\text{const}$, Eq.(\ref{mub}) reduces to $\vec{n_a}.\vec{n_b}=0$, i.e. the Bloch vectors corresponding to mutually unbiased state vectors are orthogonal to each other,  which is similar to mutually unbiased qubits. Here we require that for orthogonal Bloch vectors the angular parameters obey the relation $\cos(\theta_{ai}-\theta_{bi})=\text{const}$. For $\cos(\theta_{ai}-\theta_{bi})\neq \text{const}$, the condition is more complex.

\section{Implications of qutrit Bloch sphere construction}\label{VI}
\subsection{Employing the Bloch sphere geometry to find MUBs in three dimensions}
It is known that in prime or power of prime dimension $d=p^n$, where $p$ is a prime number and $n$ is an integer greater than zero, there exist a maximum of $d+1$ MUBs \cite{Bandyopadhyay2002}. For the qubit, the existence of three MUBs can be very easily explained through the qubit Bloch sphere, but such an explanation is difficult in higher-level quantum systems. In this section, we show that the qutrit Bloch sphere geometry restricts the maximum number of MUBs to four.

\textit{MUBs in $2$ dimensions}- The qubit Bloch sphere is a three-dimensional sphere, in which the Bloch vectors corresponding to orthonormal basis kets lie on the antipodal points on the sphere, i.e. they lie along the line passing through the center. Also, the Bloch vectors corresponding to mutually unbiased kets are orthogonal to each other \cite{sharma2019fine}. As, there can be only three such orthogonal lines passing through the center, which explains why there are only three possible mutually unbiased bases in dimension 2.

\textit{MUBs in $3$ dimensions}.-- To find the qutrit MUBs, we first fix one of the orthonormal basis to be the eigenbasis of HW operator $Z$ or the computational basis. The eigenvectors of $Z$ have the following Bloch vector and angular parameters
\begin{align*}
(n_2=1,\theta_2=0)\rightarrow \ket{0},\:\: \left(n_2=-1,\theta_2=\frac{\pi}{3}\right)\rightarrow\ket{1},\:\: {\rm and}\:\: \left(n_2=1,\theta_2=\frac{2\pi}{3}\right)\rightarrow\ket{2}.
\end{align*}
Notice, here that the pairs $(n_2,\theta_2)$ in finding computational basis are not only choices, but they are also one of the possible combinations (see Eq. \ref{rank-1}). 

According to Eq.(\ref{mub}), any pure qutrit which is mutually unbiased to all the computational basis must have $n_2=0$. However, finding such pure states are straight forward as is seen from Eq. (\ref{rank-1}). We list the other three MUBs below,  
%and $\theta_2$ can have any arbitrary value, which can be also deduced from Eq.(\ref{mub}). Hence, we need to find the remaining mutually unbiased kets using the remaining weight parameters $n_1,n_3$ \text{and} $n_4$ and the angular parameters $\theta_1,\theta_3,\theta_4$. 
\begin{align*}
(n_k=1,\theta_k=0)\rightarrow \ket{+}_k,\:\: \left(n_k=-1,\theta_k=\frac{\pi}{3}\right)\rightarrow\ket{-}_k,\:\: {\rm and}\:\: \left(n_k=1,\theta_k=\frac{2\pi}{3}\right)\rightarrow\ket{\omega}_k,
\end{align*}
where $k=1,3,4$. One can easily find their expressions by putting these values in the general expression of the qutrit density matrix. 

\subsection{Characterization of Unital Maps}
In this section, we characterize the unital maps acting on the qutrit states. Unital maps are quantum operations that preserve the identity matrix or the maximally mixed density matrix. It is known that the unital maps acting on a qubit density matrix are characterized by a convex tetrahedron\cite{904522,BETHRUSKAI2002159}. 

To analyze the unital channels acting on a qutrit density matrix $\rho=\frac{1}{3}\sum_{p,q}^{2}b_{pq}U_{pq}$ with Bloch vector $\vec{b}_{pq}$ (see Eq.(\ref{weylexp})), we note that a linear quantum map can be written in the form of an affine transformation acting on the $d^2-1=8$ dimensional Bloch vector. Thus, every linear qutrit quantum map $\Phi:\mathbb{C}^{3\times 3}\rightarrow\mathbb{C}^{3\time 3}$ can be represented using a $9\times 9$ matrix $\mathcal{L}$ acting on the column vector $\{1, \vec{b}_{pq}\}$. The action of the quantum channel $\rho\rightarrow \Phi(\rho)=\frac{1}{3}\sum_{p,q}^{2}b_{pq}'U_{pq}$ can be written as 
\begin{align*}
\vec{b}\rightarrow\vec{b'}=L\vec{b}+\vec{l},\:\:{\rm with}\:\: \mathcal{L}=\begin{pmatrix}
1&0\\l&L \end{pmatrix},  
\end{align*}
where $L$ is an $8\times 8$ matrix and $l$ is a column vector containing eight elements. 
By observing Eq.(\ref{Blochrelations}), it can be seen that to make sure that $\vec{b'}$ corresponds to a hermitian density matrix $\mathcal{E}(\rho)$, it is necessary that 1) $L$ is a diagonal matrix with eigenvalues \{$\lambda_{01},\lambda_{02},...,\lambda_{22} $\}  and 2) the eigenvalues must be of the following form 
\begin{align*}
&\lambda_{01}=\lambda_1 e^{i\phi_1}, \lambda_{02}=\lambda_1 e^{-i\phi_1},\:\: \lambda_{10}=\lambda_2 e^{i\phi_2}, \lambda_{20}=\lambda_2 e^{-i\phi_2}\nonumber\\
&\lambda_{12}=\lambda_3 e^{i\phi_3}, \lambda_{21}=\lambda_3 e^{-i\phi_3},\:\: \lambda_{22}=\lambda_4 e^{i\phi_4}, \lambda_{11}=\lambda_4 e^{-i\phi_4}.
\end{align*}
Next, we note that to preserve the identity matrix, $\vec{l}=\vec{0}$. Now, to do the complete characterization of the  map $\mathcal{L}$ we impose the complete positivity requirement via Choi's theorem which requires that Choi Matrix $\mathbf{C}=(\mathbb{I}\otimes\Phi)(|\Omega\rangle\langle\Omega|)$ is positive semidefinite, $|\Omega\rangle=\sum_{i}|ii\rangle$. 
To simplify the problem, we find the eigenvalues when the angles $\phi_i=0$. The constraints on the parameters \{$\lambda_i$\}'s are given by
\begin{align}\label{unitalpolygon}
&1+2\lambda_1-\lambda_2-\lambda_3-\lambda_4\geq 0,\:\: 1-\lambda_1+2\lambda_2-\lambda_3-\lambda_4\geq 0,\:\: 1-\lambda_1-\lambda_2+2\lambda_3-\lambda_4\geq 0,\nonumber\\
&1-\lambda_1-\lambda_2-\lambda_3+2\lambda_4\geq 0,\:\: {\rm and}\:\:1+2\lambda_1+2\lambda_2+2\lambda_3+2\lambda_4\geq 0.
\end{align}
One can easily generalize this result for arbitrary $\phi$ values by noticing that the change in $H_\ell$'s are happening by $e^{\mi \theta}U_{pq}+e^{-\mi \theta}U_{-p,-q}\rightarrow e^{\mi (\theta+\phi)}U_{pq}+e^{-\mi (\theta+\phi)}U_{-p,-q}$ which preserves the Hermiticity and trace orthogonality of $H_\ell$. 
The above constraint gives a convex polygon space with five vertices 
\begin{align*}
v_1=\{\bm 1\},\:\: v_2=\{1,-\frac{\bm 1}{\bm 2}\}, \:\:
v_3=\{-\frac{1}{2},1,-\frac{1}{2},-\frac{1}{2}\},\:\: v_4=\{-\frac{1}{2},-\frac{1}{2},1,-\frac{1}{2}\},\:\: {\rm and}\:\:
v_5=\{-\frac{\bm 1}{\bm 2},1\}.
\end{align*}
It is an irregular polygon with 8 edges, out of which 4 edges have Euclidean length $\sqrt{9/2}$ and 4 other edges have Euclidean length $\sqrt{27/4}$. 

It is insightful to visualize the effect of the action of the channel on a state in the sphere $\mathbb{R}^4$. The parameters \{$\lambda_i$\} reduce the length of each Bloch vector component from $n_i$ to $\lambda_in_i$, thus bringing the state closer to the origin. 
%Now notice that we have obtained the constraints in Eq.(\ref{unitalpolygon}) assuming that all $\phi_i=0$, i.e., there is no change in the angular parameters $\theta_i$. The allowed values of $\phi_i$'s will therefore depend on how much the lengths of the Bloch vector components have been reduced by $\lambda_i$. 

%Instead of Weyl operators if we use the Gell-Mann operators in this analysis to obtain the constraints like Eq.\ref{unitalpolygon}. It does not give a convex polytope like structure. Hence, the characterization of unital maps is not possible.

\subsection{Characterization of Randomly Generated Density Matrices}
In this section, we characterize the structure of the state space of randomly generated density matrices, using the Bloch sphere in $\mathbb{R}^4$. Specifically, we show the representation of ensembles generated by Hilbert-Schmidt and Bures metrics \cite{zyczkowski2011generating,Zyczkowski_2001}. 
The infinitesimal Hilbert-Schmidt (Eq.(\ref{HS})) distance between $\rho$ and $\delta\rho$, has a very simple form given as $d_{HS}^2={\rm Tr}[(\delta\rho)^2]$. In $n$-dimensions, the probability distribution induced by this metric, derived by Hall \cite{HALL1998123} is given by 
\begin{align}\label{HSprob}
	P_{HS}(\lambda_1,...,\lambda_n)=C_{HS}\delta\left(1-\sum_{i=1}^n\lambda_i\right)\prod_{j<k}^{n}(\lambda_j-\lambda_k)^2,
\end{align}
where $\lambda_i$'s are the eigenvalues of $\rho$ and $C_{HS}$ is determined by the normalization. 

For mixed quantum states, there is another useful distance measure known as the Bures distance\cite{bures1969extension,uhlmann1992metric}
\begin{align*}
D_{B}^2(\rho_1,\rho_2)=2\left(1-{\rm Tr}\left[\sqrt{\sqrt{\rho_1}\rho_2\sqrt{\rho_1}}\right]\right).
\end{align*}
Similar to the Hilbert-Schmidt case, there exists the infinitesimal form Bures metric derived by Hubner \cite{hubner1992explicit} 
\begin{align*}
d_{B}^2=\frac{1}{2}\sum_{j,k=1}^{n}\frac{|\bra{j}\delta\rho\ket{k}|^2}{\lambda_j+\lambda_k},
\end{align*}
where again $\lambda_k$ and $\ket{k}$ are respectively the eigenvalues and eigenvectors of $\rho$. For this metric also, the probability distribution was derived by Hall \cite{HALL1998123}, which is given by 
\begin{align}\label{Buresprob}
		P_{B}(\lambda_1,...,\lambda_n)=C_{B}\frac{\delta(1-\sum_{i=1}^n\lambda_i)}{(\lambda_1\cdot\lambda_1\cdots\lambda_n)^{1/2}}\prod_{j<k}^{n}\frac{(\lambda_j-\lambda_k)^2}{\lambda_j+\lambda_k},
\end{align}
where $C_{B}$ is again determined by the normalization.
In Eqs. (\ref{HSprob}) and (\ref{Buresprob}), we have the probability distributions defined on the simplex of eigenvalues. However, we want to see how this probability distribution picks out the states from the Bloch sphere. For a two-dimensional state $\rho=(1/2)(\mathbb{I}+\vec{r}\cdot\sigma)$, we can translate the eigenvalues to Bloch sphere parameters using the simple formulas $\lambda_1=(1+r)/2$ and $\lambda_2=(1-r)/2$, where $\lambda_1, \lambda_2$ are the two eigenvalues of $\rho$. By substituting these in Eqs.(\ref{HSprob}) and (\ref{Buresprob}), we get the following probability distributions in terms of Bloch sphere parameters \cite{HALL1998123}

%\begin{align*}
%	&P_{HS}(r)=6r^2, \\
%	&P_{Bures}(r)=\frac{8r^2}{\pi\sqrt{1-r^2}}.
%\end{align*}
%The radial form of these distributions takes the following form \cite{HALL1998123} 
\begin{align}
P_{HS}(\vec{r})=\frac{3}{4\pi},\:\:{\rm and}\:\:P_{B}(\vec{r})=\frac{4}{\pi\sqrt{1-r^2}}.\label{qubitBures}
\end{align}
We can see that both probability distributions are dependent only on the radial parameter $r$. While the HS distribution is uniform over the Bloch sphere while the Bures distribution is sharply peaked at the surface of the Bloch sphere.

Next, we derive the form of these probability distributions with respect to our representation of qutrit states. For a qutrit state $\rho$, its eigenvalues $\lambda_1,\lambda_2$ and $\lambda_3$ can be written directly in terms of the Bloch sphere parameters $n_i$'s and angular parameters $\theta_i$. However, a direct approach will lead to cumbersome calculations. Instead, we write the eigenvalues $\lambda_i$'s in terms of the characteristic equation coefficients $a_i$'s from Eq.(\ref{chareq}) and substitute in the Eqns.(\ref{HSprob}) and (\ref{Buresprob}) which gives us the following 
%\begin{align*}
%	&P_{HS}(r,\zeta_i,\theta_i)\nonumber \\
%	&=\frac{\left((r-1)^2(2r+1)-27 \text{Det}(\rho)\right)\left((r+1)^2(2r-1)+27 \text{Det}(\rho)\right)}{27}, \\
%    &P_{Bures}(r,\zeta_i,\theta_i)\nonumber \\
%    &=\frac{\left((r-1)^2(2r+1)-27 \text{Det}(\rho)\right)\left((r+1)^2(2r-1)+27 \text{Det}(\rho)\right)}{27\left(\frac{1-r^2}{3}-\text{Det}(\rho)\right)\text{Det}(\rho)}.
%\end{align*}
\begin{align}
	P_{HS}(\vec{r},\alpha_i,\theta_i)=\frac{C_{HS}F(\rho)}{27r^3},
 %\label{qutritHSHW} 
 \:\:{\rm and}\:\:P_{B}(\vec{r},\zeta_i,\theta_i)=\frac{C_{B}F(\rho)}{27r^3 \left(\frac{1-r^2}{3}-\text{Det}(\rho)\right)\sqrt{\text{Det}(\rho)}}.\label{qutritBuresHW}
\end{align}
where $F(\rho)=\left\{(r-1)^2(2r+1)-27 \text{Det}(\rho)\right\}\left\{(r+1)^2(2r-1)+27 \text{Det}(\rho)\right\}$ and we have switched to the polar representation with  $n_1=r\cos\alpha_1,n_2=r\sin\alpha_1\cos\alpha_2,n_3=r\sin\alpha_1\sin\alpha_2\cos\alpha_3$ and $n_4=r\sin\alpha_1\sin\alpha_2\sin\alpha_3$. Also, $C_{HS}$ and $C_{B}$ are constants and are determined by normalization. In this form, these probability distributions don't give much information about the states in the Bloch sphere because of dependence on the angular parameters $\theta_i$'s which are not a part of the sphere in $\mathbb{R}^4$. We can obtain a distribution for a subset of states by fixing the $\theta_i$'s and then analyze the probability distributions. After fixing all the $\theta_i$'s values(say all zero), we get Det$(\rho)=f(r,\bm\alpha)$. The distributions in Eq.(\ref{qutritBuresHW}) are not invariant with respect to unitary operations unlike in the qubit scenario. This is a signature of the fact that all points inside the Bloch sphere in $\mathbb{R}^4$ don't represent physical states.

After some algebraic calculations, it is found that the HS distribution in Eq.(\ref{qutritBuresHW}) is always positive irrespective of Det($\rho)$ being positive or negative. Whereas the Bures distribution in Eq.(\ref{qutritBuresHW}) is positive if and only if Det($\rho)\geq0$, hence picking out the closed structure of the qutrit states inside the Bloch sphere. Moreover, the HS distribution is non-decreasing with respect to the radial parameter $r$, everywhere. Whereas, the Bures distribution is non-decreasing with respect to $r$ in the region where the Det($\rho)\geq0$. It can also be seen that the Bures distribution is sharply peaked whenever the denominator vanishes. While $\text{Det}(\rho)=0$ for rank-2 or rank-1 states, the $\{(1-r^2)/3\}-\text{Det}(\rho)=0$ only at the surface of the Bloch sphere or beyond.

Thus if we fix the $\theta_i$'s, both these distributions are localized closer to the  surface of the Bloch sphere. For the HS distribution, this is unlike what happens in the qubit scenario where it is uniform all over the sphere. Whereas, the Bures distribution is sharply peaked near or at the surface of the Bloch sphere. It is similar to the behavior of the Bures distribution in the qubit scenario, where the Bures distribution is sharply peaked on the surface. These results are matching with the plots presented in Fig.2 of Ref.\cite{Zyczkowski_2001}, which depicts the plots in the simplex of eigenvalues.

As an example, we fix the all $\theta_i=0$' and all polar angles $\alpha_i$'s as $\alpha_1=\pi/3,\alpha_2=0,\alpha_3=\pi/7$, to see the dependence on the radial parameter $r$, and obtain the following 
\begin{align}
	P_{HS}(r)=C_{HS}\frac{6-\sqrt{3}}{72}r^3,\:\:P_{B}(r)=C_{B}\frac{162(6-\sqrt{3})r^3}{(\sqrt{4-12r^2+6.19r^3})(-32+24r^2+6.19r^3)}.
\end{align}
We see that in the chosen direction, HS distribution is peaked on the surface of the Bloch sphere and it is everywhere positive. While the Bures distribution sharply peaked at $r \approx0.73$ and while is negative for $r>0.73$. It simply tells that for the chosen $\theta_i$'s there are no more physical states beyond $r\approx0.73$ in the chosen direction and also that there is a rank 2 state at $r\approx0.73$. The other singularity of the Bures distribution lies at $r \approx 1.02$, but $P_B(r)$ is negative after $r=0.73$ and hence we ignore it.

In \ref{Gell-Mann}, we also do the analysis of HS and Bures distributions when the qutrit states are represented using Gell-Mann operators. There also we make similar observations, i.e., 1) The HS distribution is always positive whereas the Bures distribution is positive iff Det$(\rho)\geq 0$. 2) HS distribution is non-decreasing with respect to the radial parameter and hence the states are localized on the surface of the convex structure of the states and 3) Bures distribution is non-decreasing for Det$(\rho)\geq 0$ and it also blows up at the surface of the Bloch sphere or for the rank-2 states.

%\section{Bloch vector for qudits}\label{VII}
%This approach of separating the weight parameters and angular parameters, in the HW operator-based representation can also be extended to higher dimensions. If we consider a qudit of dimension $d$, the number of HW operators is $d^2-1$ (excluding the identity matrix), among them there will be $d+1$ sets of commuting HW operators containing $d-1$ HW operators each. Each set will have one weight parameter associated with it as $w_i$ and $d-2$ angular parameters so that there are total $d^2-1$ real parameters. In this way, one can create a $d+1$ dimensional Bloch sphere built from $d+1$ weight parameters. However, as dimensions increase this analysis will only become more complex. 
\section{Extention to $d\geq 4$}\label{VII}
%\section{Forming new basis from HW basis}
In this section, we extend the above analysis to $d\geq 4$. Our aim is to find the dimension of Bloch sphere geometry in these dimensions. We find that 1) for prime $d$, the Bloch sphere lives in $\mathbb{R}^{(d^2-1)/2}$, however, 2) for non-prime $d$, it is hard to tell precisely. 

We find that it is possible to find such a group of Hermitian matrices from the HW basis. Below, we describe our method in detail. Our aim is to find two properties of HW operators $\{U_{pq}\}$, namely, 
\begin{itemize}
    \item the conditions that pairs of HW matrices are complex-conjugate to each other,
    \item the conditions that some HW matrices are forming a coset of pairwise commuting matrices.
\end{itemize}
To find the complex-conjugate of $U_{pq}$, we recall the relation that $U_{pq}^\dagger=\omega^{pq}U_{-p,-q}$. This means that the HW matrix $U_{\ell,m}$, which will be equal to $U_{pq}^\dagger$, should satisfy the relation that $\ell+p=m+q=nd$, where $n=0,...,d-1$. Clearly, it is always possible to find complex conjugates of $U_{pq}$ within the set of HW matrices, $\{U_{pq}\}$. Now, let us consider that the coset $\{U_{\ell m}\}$ that are mutually commuting, then they should satisfy the following property 
\begin{align*}
    0&=U_{\ell m}U_{\ell' m'}-U_{\ell' m'}U_{\ell m}\\
    &=\omega^{\frac{\ell m+\ell' m}{2}}X^{m+m'}Z^{\ell+\ell'}(\omega^{\ell m'}-\omega^{\ell' m}).
\end{align*}
Therefore, for mutual commutavity, $\ell m'=\pm \ell' m$, where $\pm$ is modulo $d$. This condition can compactly be written as $\ell m=nd+k\leq (d-1)^2$, where $k=0,...,d-1$. The last inequality comes from the fact that both $(\ell,m)$ can have maximum value $d-1$. How many such cosets exist? If we count the possible $k$ values, the number of cosets are always $d+1$, as `$k=0$' can come from two distinct possibilities $(\ell,m)=(0,q)$ and $(p,0)$. However, note that $k=0$ can come from $\ell m=nd$ also, and we hoped that these elements might be distributed inside one of the cosets $\{(0,q)|q=1,..,d-1\}$ and $\{(p,0)|p=0,...,d-1\}$ depending with which coset they commute. However, we find that this is never the case in general for tractable dimensions. Therefore, we ask: How many elements exist in each coset? Naturally, the answer to this isn't straightforward. We will answer this question in the following sections.

\subsection{$\mathbb{R}^{(d^2-1)/2}$ Bloch sphere representation for qudits with prime $d$}
For prime power dimensions, below, we state a known result in the literature: 
\bq
{\bf Claim.1}: There exists $d+1$ cosets consisting of $d-1$ mutually commuting HW matrices.
\eq
Along with the above claim, we observe that 
\bq
{\bf Corollary}: Within each cosets --
\begin{enumerate}
    \item for all prime $d$, individual cosets contains the pairs $\{U_{\ell m},U_{\ell' m'}\}$ which are each others complex-conjugate. That means $(d-1)/2$ such pairs exist in a coset.
    \item for prime $d$, there exists no $U_{\ell m}$ such that $U_{\ell m}^\dagger=U_{\ell m}$.
\end{enumerate}
\eq
Now if we recall Eq.(\ref{weylexp}), and apply the above properties, we can conclude that for pair of commuting HW matrices, $U_{\ell m}^\dagger=U_{\ell' m'}$ such that $b_{\ell m}=n_ie^{\mi \theta_i}=b_{\ell' m'}^*$, where $(n_i,\theta_i)\in \mathbb{R}$. This means one finds terms like $b_{\ell m}U_{\ell m}+b_{\ell' m'}U_{\ell' m'}$ inside $\rho$, which can be rewritten as $n_i H_i$, where $H_i=e^{\mi \theta_i}U_{\ell m}+e^{-\mi \theta_i}U_{\ell' m'}$. Note that all $\{H_i\}$ satisfy both $H_i^\dagger=H_i$, and ${\rm Tr}[H_i^\dagger H_j]=2d\delta_{ij}$. 
%Moreover, within each cosets all such $H_i$'s will commute with each other, therefore, one will find $\Tilde{H}_i$ with properties $\Tilde{H}_i^\dagger=\Tilde{H}_i$ and ${\rm Tr}[\Tilde{H}_i^\dagger \Tilde{H}_j]=d\delta_{i,j}$ such that 
%\begin{align*}
  %  \Tilde{H}_i=\sum_{j=1}^{(d-1)/2}(n_j/n)H_j,\:\: {\rm where}\:\: n^2=\sum_{j=1}^{(d-1)/2}n_j^2.
%\end{align*} 
This means, for prime $d$, we find a group of $(d^2-1)/2$ trace-orthogonal and Hermitian matrices $\{H_i\}$. 
%This group is called abelian sub-group (\textcolor{red}{!Caution: Is it under HS norm for matrices?}). 
Notice that these matrices are no longer unitary. Then any density matrices in prime $d$ can be written as 
\begin{align}\label{d+1Dstate}
    \rho=\frac{1}{d}\left[\mathbb{I}+\bm n .\bm H\right],\:\: {\rm with}\:\: n_k=\frac{1}{2}{\rm Tr}[\rho H_k],
\end{align}
where $\bm n$ is a $(d^2-1)/2$-dimensional real vectors with $|\bm n|^2\leq (d-1)/2$. We call $n_i$s the weight parameters. It should also be noted that the angular parameters ($\theta_i$s) can be estimated by the following formula
\begin{align}
    \theta_i=\arccos\left[\frac{1}{n_i}{\rm Re}\Big({\rm Tr}[\rho U_{\ell m}]\Big)\right].
\end{align}
{\bf Comment}: -- Our construction is inducing a Bloch sphere in $\mathbb{R}^{(d^2-1)/2}$. Effectively, we are reducing in terms of the dimension of Euclidean space. However, we are having $(d^2-1)/2$ number of $\theta$ parameters which induces an envelope in the state-space dictating valid regions. 
\subsubsection{Bloch sphere representation of a state in $d=5$}\label{IIIb}
For the states in $d=5$, there are six possible cosets; $\{U_{p0}|p=1,..,4\}$; $\{U_{0p}|p=1,...,4\}$; $\{U_{11},U_{23},U_{32},U_{44}\}$; $\{U_{12},U_{21},U_{34},U_{43}\}$; $\{U_{13},U_{24},U_{31},U_{42}\}$; and $\{U_{14},U_{22},U_{33},U_{41}\}$.
%one can find the HW basis in a following way $U_{pq}=\{U_{00}=\mathbb{I},U_{0q}=X^q,U_{p0}=Z^p,U_{pq}=\prod_q(XZ^p)^q| (q,p)=1,..,4\}$, where $X,Z$ are defined in Eq. (\ref{XandZ}). 
Using the same analysis from the previous subsection, we find that we can consider a set of Hermitian, traceless, trace-orthogonal matrices $\{H_i\}$ of the form 
\begin{align*}
   & H_1=e^{\mi\theta_1}U_{10}+e^{-\mi\theta_1}U_{40},\:\: H_{2}=e^{\mi\theta_{2}}U_{20}+e^{-\mi\theta_{2}}U_{30},\:\: H_{3}=e^{\mi\theta_{3}}U_{01}+e^{-\mi\theta_{3}}U_{04},\\ 
   &H_{4}=e^{\mi\theta_{4}}U_{02}+e^{-\mi\theta_{4}}U_{03},\:\: H_{5}=e^{\mi\theta_{5}}U_{11}+e^{-\mi\theta_{5}}U_{44},\:\: H_{6}=e^{\mi\theta_{6}}U_{23}+e^{-\mi\theta_{6}}U_{32}, \\ 
   &H_{7}=e^{\mi\theta_{7}}U_{12}+e^{-\mi\theta_{7}}U_{43},\:\: H_{8}=e^{\mi\theta_{8}}U_{21}+e^{-\mi\theta_{8}}U_{34},\:\: H_{9}=e^{\mi\theta_{9}}U_{13}+e^{-\mi\theta_{9}}U_{42}, \\ 
   &H_{10}=e^{\mi\theta_{10}}U_{24}+e^{-\mi\theta_{10}}U_{31},\:\: H_{11}=e^{\mi\theta_{11}}U_{14}+e^{-\mi\theta_{11}}U_{41},\:\: H_{12}=e^{\mi\theta_{12}}U_{22}+e^{-\mi\theta_{12}}U_{33}, 
\end{align*}
where $\theta_i\in \mathbb{R}$. Therefore, one can write the state in $d=5$ as 
\begin{align}\label{5-state}
\rho=\frac{1}{5}\left[\mathbb{I}+\bm n .\bm H\right],\:\: {\rm with}\:\: n_i=\frac{1}{2}{\rm Tr}[\rho H_i],
\end{align}
where $\bm n$ is a real vector in $\mathbb{R}^{12}$. Note here that like qutrit, these six cosets are related to six MUBs. 
%with $|\bm n|^2\leq \frac{4}{5}$. 
%Clearly, the elements of vector $\bm n$ are determined by the following relation
%\begin{align}
%    n_m=\frac{1}{2}{\rm Tr}[\rho H_m].
%\end{align}
%Now, we notice that the pair, $\{H_j\}$ within same coset, are commuting also. This implies that these pairs can be reduced to single matrix by appropriate normalization. For example, consider $H_1,H_2$, then $(1/\sqrt{n_1^2+n_2^2}) (n_1 H_1+n_2H_2)=\tilde{H}_1$, where $\tilde{H}_1$ is Hermitian. By this logic, we have $6$ such matrices $\{\tilde{H}_k\}$ with ${\rm Tr[\tilde{H}_k\tilde{H}_\ell]}=d\delta_{k\ell}$ and we can rewrite Eq.(\ref{5-state}) as
%\begin{align}\label{5-state12}
%\rho=\frac{1}{5}\left[\mathbb{I}+2\bm \tilde{n} .\bm \tilde{H}\right],
%\end{align}
%where $\bm \tilde{n}$ is a $6$-$D$ real vectors. 
%with $|\bm \tilde{n}|^2\leq \frac{4}{5}$. 
%Clearly, the elements of vector $\bm n$ are determined by the following relation
%\begin{align}
   % \tilde{n}_k=\frac{1}{2}{\rm Tr}[\rho \tilde{H}_k].
%\end{align}
%%%%%%%%%%%%%%%%%%%%%%%%%%%%%%%%%%%%%%%%%%%%%%
\subsection{Bloch sphere representation of qudit when $d$ is non-prime}\label{IVn}
\bq
{\bf Claim.2}: There exists $d+1$ such cosets of HW matrices plus some extra cosets from the relation $\ell m=nd+0$ whenever $\ell$ or $m\neq 0$.
\eq
\bq
{\bf Corollary}: Within such cosets --
\begin{enumerate}
    \item for all non-prime $d$, individual cosets contains the pairs $\{U_{\ell m},U_{\ell' m'}\}$ which are each others complex-conjugate. 
    \item for non-prime $d$, there exist at most three $U_{\ell m}$ such that $U_{\ell m}^\dagger=U_{\ell m}$ and they are $(\ell,m)=\{(d/2,0),(0,d/2),(d/2,d/2)\}$. For some non-prime $d$, there exists none, eg., $d=9,25,27, ...$etc.
    \item a coset can contain at least one HW matrix.
\end{enumerate}
\eq
Notice that arbitrary density matrix in non-prime $d$ will also be concisely written as Eq.(\ref{d+1Dstate}), however, the dimension of the Bloch vector is not precisely known as is shown in the below examples.
\subsubsection{d=4}
There are five (\sout{six}) possible cosets for $d=4$ and they are listed in Table \ref{d=4coset}. In this case, there are exactly three Hermitian HW matrices, $U_{02},U_{20}$, and $U_{22}$. Using the property of density matrix, $\rho^\dagger=\rho$, we find that there exist Hermitian, trace-orthogonal matrices $\{H_i\}$, with self-adjoint ones
\begin{align*}
    G1:\:\: H_2=U_{20},\:\: H_4=U_{02},\:\: H_6=U_{22},
\end{align*}
%where $\tN_i,n_i,\theta_i\in \mathbb{R}$. 
where ${\rm Tr}[H_iH_j]=4\delta_{ij}$ for $H_i\in G1$, and the expression for other $H_i$'s are defined as
\begin{align*}
    &G2:\:\:H_1=e^{\mi\theta_1}U_{10}+e^{-\mi\theta_1}U_{30},\:\: H_3=e^{\mi\theta_3}U_{01}+e^{-\mi\theta_3}U_{03}, \:\:
    H_5=e^{\mi\theta_5}U_{11}+e^{-\mi\theta_5}U_{33}, \\
    &H_7=e^{\mi\theta_7}U_{12}+e^{-\mi\theta_7}\omega^2U_{32},\:\: H_8=e^{\mi\theta_8}U_{21}+e^{-\mi\theta_8}\omega^2U_{23}, \:\:H_9=e^{\mi\theta_9}U_{13}+e^{-\mi\theta_9}U_{31},
\end{align*}
where $n_i,\theta_i\in \mathbb{R}$ and ${\rm Tr}[H_iH_j]=8\delta_{ij}$ for $H_i\in G2$. 
Note that we multiplied $\omega^2$ in front of $U_{23}$ and $U_{32}$ to get the desired properties. Notice also that there are only six angular parameters, $\theta_i$. Therefore, the state in $d=4$ can be expressed as 
\begin{align}\label{4-state}
\rho=\frac{1}{4}\left[\mathbb{I}+\bm n .\bm H\right],\:\: n_i=\frac{1}{2^{1-f(H_i)}}{\rm Tr}[H_i\rho],
\end{align}
where $f(H_i)=1$ if $H_i\in G1$, otherwise $0$, and $\bm n$ is a real vector in $\mathbb{R}^{9}$ with $\sum_i2^{1-f(H_i)}|n_i|^2\leq 3$.
%are $\{(k,0|k=1,2,3)\}$, $\{(0,k|k=1,2,3)\}$, $\{(2,2)\}$, $\{(1,2),(2,1),(3,2),(2,3)\}$, $\{(1,1),(3,3)\}$, and $\{(1,3),(3,1)\}$. One has $6$ cosets.

\begin{table}[H]
\centering
\begin{tabular}{ |l|l|l| }
\hline
 \: & \multicolumn{2}{ |c| }{Cosets ($d=4$)} \\
\hline
$k$ & \multicolumn{1}{ |c| }{$k$} & $nd+k$ \\ \hline
$0$ & $\{(p,0)\}$; $\{(0,p)\}$ & \sout{$\{(2,2)\}$} \\
\hline
$1$ & \multicolumn{2}{ |c| }{$\{(1,1),\textcolor{red}{(2,2)},(3,3)\}$} \\
\hline
$2$ & \multicolumn{2}{ |c| }{$\{(1,2),(2,1),(2,3),(3,2)\}$} \\ \hline
$3$ & \multicolumn{2}{ |c| }{$\{(1,3),(3,1)\}$} \\
\hline
\end{tabular}
\caption{\textcolor{blue}{Coset for $d=4$}.- Here $p=1,2,3$. Notice that for $k=0$ there exists three distinct cosets. The lone coset $\{(2,2)\}$ is compatible with the coset for $k=1$, forming a perfect coset with $d-1$ elements. This indicates that we might find at least $3$ MUBs.}
\label{d=4coset}
\end{table}

\subsubsection{d=6}
In $d=6$, a total of nine cosets exist and they are listed in Table \ref{d=6coset}. Here also, we find that exactly three Hermitian HW matrices exist, which are  $U_{03},U_{30}$, and $U_{33}$. Notice that the the cosets $\{(2,3),(4,3)\}$ and $\{(3,2),(3,4)\}$ do not commute with any other cosets from Table \ref{d=6coset}.
%, indicating that we need atleast nine real parameters to represent the state in $d=6$. 
By a similar argument, we find that there exist Hermitian, trace-orthogonal matrices $\{H_i\}$, with 
\begin{align*}
    G1:\:\:H_3=U_{30},\:\: H_6=U_{03},\:\: H_9=U_{33},
\end{align*}
where ${\rm Tr}[H_iH_j]=6\delta_{ij}$ for $H_i\in G1$, and the expression for other $H_i$'s are defined as
\begin{align*}
    &G2:\:\: H_1=e^{\mi\theta_1}U_{10}+e^{-\mi\theta_1}U_{50}, H_2= e^{\mi\theta_2}U_{20}+e^{-\mi\theta_2}U_{40}, H_4=e^{\mi\theta_4}U_{01}+e^{-\mi\theta_4}U_{05},\\
    &H_5=e^{\mi\theta_5}U_{02}+e^{-\mi\theta_5}U_{04},
    H_7=e^{\mi\theta_7}U_{13}+e^{-\mi\theta_7}U_{53}, H_8=e^{\mi\theta_{8}}U_{31}+e^{-\mi\theta_{8}}U_{35}, \\
    &H_{10}=e^{\mi\theta_{10}}U_{12}+e^{-\mi\theta_{10}}\omega^3U_{54}, H_{11}=e^{\mi\theta_{11}}U_{21}+e^{-\mi\theta_{11}}\omega^3U_{45}, H_{12}=e^{\mi\theta_{12}}U_{24}+e^{-\mi\theta_{12}}U_{42},\\
    & H_{13}=e^{\mi\theta_{13}}U_{11}+e^{-\mi\theta_{13}}U_{55},H_{14}=e^{\mi\theta_{14}}U_{15}+e^{-\mi\theta_{14}}U_{51},H_{15}=e^{\mi\theta_{15}}U_{23}+e^{-\mi\theta_{15}}\omega^3U_{43},\nonumber \\
    &H_{16}=e^{\mi\theta_{16}}U_{14}+e^{-\mi\theta_{16}}\omega^3U_{52}, H_{17}=e^{\mi\theta_{17}}U_{41}+e^{-\mi\theta_{17}}\omega^3U_{25}, H_{18}=e^{\mi\theta_{18}}U_{22}+e^{-\mi\theta_{18}}U_{44},
    \\
    & H_{19}=e^{\mi\theta_{19}}U_{32}+e^{-\mi\theta_{19}}\omega^3U_{34},
\end{align*}
where $\theta_i\in \mathbb{R}$ and ${\rm Tr}[H_iH_j]=12\delta_{ij}$ for $H_i\in G2$. Note that we multiplied $\omega^3$ in front of certain HW matrices to get the desired properties. Therefore, the density matrix in $d=6$ can be written as 
\begin{align}\label{6-state}
\rho=\frac{1}{6}\left[\mathbb{I}+\bm n .\bm H\right],\:\: n_i=\frac{1}{2^{1-f(H_i)}}{\rm Tr}[H_i\rho],
\end{align}
where $f(H_i)=1$ if $H_i\in G1$, otherwise $0$, and $\bm n$ is a real vector in $\mathbb{R}^{19}$ with $\sum_i2^{1-f(H_i)}|n_i|^2\leq 5$.
%$\{(k,0|k=1,...,5)\}$, $\{(0,k|k=1,...,5)\}$, $\{(2,3),(4,3)\}$, $\{(3,2),(3,4)\}$, $\{(1,1),(5,5)\}$,\\ $\{(1,2),(2,1),(2,4),(4,2),(4,5),(5,4)\}$, $\{(1,3),(3,1),(3,3),(3,5),(5,3)\}$, $\{(1,4),(4,1), (2,2),(2,5),(5,2), (4,4)\}$, and $\{(1,5),(5,1)\}$. It has $9$ cosets.
\begin{table}[H]
\centering
\begin{tabular}{ |l|l|l| }
\hline
 \: & \multicolumn{2}{ |c| }{Cosets ($d=6$)} \\
\hline
$k$ & \multicolumn{1}{ |c| }{$k$} & \multicolumn{1}{ |c| }{$nd+k$} \\ \hline
$0$ & $\{(p,0)\}$ ; $\{(0,p)\}$ & $\{(2,3),(4,3)\};\{(3,2),(3,4)\}$ \\
\hline
$1$ & \multicolumn{2}{ |c| }{$\{(1,1),(5,5)\}$} \\
\hline
$2$ & \multicolumn{2}{ |c| }{$\{(1,2),(2,1),(2,4),(4,2),(4,5),(5,4)\}$} \\ \hline
$3$ & \multicolumn{2}{ |c| }{$\{(1,3),(3,1),(3,3),(3,5),(5,3)\}$} \\
\hline
$4$ & \multicolumn{2}{ |c| }{$\{(1,4),(4,1),(2,2),(2,5),(5,2),(4,4)\}$} \\
\hline
$5$ & \multicolumn{2}{ |c| }{$\{(1,5),(5,1)\}$} \\
\hline
\end{tabular}
\caption{\textcolor{blue}{Coset for $d=6$}.- Here $p=1,..5$. Notice that for $k=0$ there exists four distinct cosets. We find that there exists $3$ perfect cosets, $k=0$ ($\times 2$), and $k=3$, indicating the existence of at least $3$ MUBs.}
\label{d=6coset}
\end{table}
\subsection{Finding MUBs in non-prime $d$}
For completeness, we extend the analysis of finding the MUBs to non-prime $d$ using our construction. Note that in every dimension, the presence of a coset with $d-1$ HW matrices might imply that there exists a MUB. 

\textit{MUBs in $4$ dimensions}.-- From Table \ref{d=4coset}, we know that it is possible to find $3$ MUBs in $d=4$. Then we have the computational basis below with the notation, $(n_1,n_2, \theta_1)\rightarrow\ket{k}$:
\begin{align*}
    (\mathcal{B}1):\:\:(1,1,0)\rightarrow\ket{0},\:\:\left(-1,-1, \frac{\pi}{2}\right)\rightarrow\ket{1}, \:\:
    (1,1, \pi)\rightarrow\ket{2},\:\:{\rm and}\:\:\left(1, -1, \frac{\pi}{2}\right)\rightarrow\ket{3}.
\end{align*}
From the other two complete cosets from Table \ref{d=4coset}, we have the two more MUBs below with the notation, $(n_3,n_4, \theta_3)\rightarrow\ket{k}$ for $\mathcal{B}2$ and $(n_4,n_5, \theta_5)\rightarrow\ket{k}$ for $\mathcal{B}3$: 
\begin{align*}
   &\mathcal{B}2:\:\: (1,1,0)\rightarrow\ket{+},\:\:\left(-1,-1, \frac{\pi}{2}\right)\rightarrow\ket{\omega}_1,\:\:\left(1,-1, \frac{\pi}{2}\right)\rightarrow\ket{\omega}_2,\:\:{\rm and}\:\:(-1, 1, 0)\rightarrow\ket{-};\\ 
   & \mathcal{B}3:\:\: (1,1,0)\rightarrow\ket{\uparrow},\:\:\left(1,-1, \frac{\pi}{2}\right)\rightarrow\ket{\Uparrow},\:\:\left(-1,-1, \frac{\pi}{2}\right)\rightarrow\ket{\Downarrow},\:\:{\rm and}\:\:(-1, 1, 0)\rightarrow\ket{\downarrow}.
\end{align*}
To find the other two MUBs using our analysis, we need to search numerically over the entire pure state space. We will pursue this in our future research.

\textit{MUBs in $6$ dimensions}.-- From Table \ref{d=6coset}, we should find 3 MUBs in $d=6$ easily. However, it is not the case. We only find two MUBs from our construction. We have the computational basis with the notation, $(n_1,n_2,n_3, \theta_1, \theta_2)\rightarrow\ket{k}$ and the other one, $\mathcal{B}2$ with the notation, $(n_4,n_5,n_6, \theta_4, \theta_5)\rightarrow\ket{k}$ below.
\begin{align*}
    &(\mathcal{B}1):\:\:(1,1,1, 0,0)\rightarrow\ket{0},\:\:\left(-1, 1,-1, 0,0\right)\rightarrow\ket{1},\:\: \left(1,-1,-1, -\frac{\pi}{3},\frac{\pi}{3}\right)\rightarrow\ket{2},\\
    &\left(-1,-1, 1, \frac{\pi}{3},-\frac{\pi}{3}\right)\rightarrow\ket{3},\:\: \left(-1,-1,1, -\frac{\pi}{3},\frac{\pi}{3}\right)\rightarrow\ket{4},\:\:\left(1,-1,-1, \frac{\pi}{3},-\frac{\pi}{3}\right)\rightarrow\ket{5}.\\
    &(\mathcal{B}2):\:\:(1,1,1, 0,0)\rightarrow\ket{+},\:\:\left(-1, 1,-1, 0,0\right)\rightarrow\ket{-},\:\:\left(1,-1,-1, -\frac{\pi}{3},\frac{\pi}{3}\right)\rightarrow\ket{\omega}_1,\\
   & \left(-1,-1, \frac{\pi}{3},-\frac{\pi}{3}\right)\rightarrow\ket{\omega}_2,\:\:\left(-1,-1,1, -\frac{\pi}{3},\frac{\pi}{3}\right)\rightarrow\ket{\omega}_3,\:\:\left(1,-1,-1, \frac{\pi}{3},-\frac{\pi}{3}\right)\rightarrow\ket{\omega}_4.
\end{align*}
Note that another MUB ($\mathcal{B}$3) can be found from any of the coset $\{(X Z^m)^p|p\in [1,5]\}$, where $m\in[1,5]$ \cite{Bandyopadhyay2002,grassl2004sicpovms,Goyeneche_2013}. Note that other properties of qudits can also be determined using Bloch parameters using our construction similar to qutrit.

\section{Discussion on the relevance of the present study with that of Ref.\cite{PhysRevA.94.010301}}\label{IX}
Before concluding, it is important for us to discuss a work (Ref. \cite{PhysRevA.94.010301}) related to our present study. The authors in Ref. \cite{PhysRevA.94.010301} consider modified HW operator basis to represent a $d$-dimensional quantum states (qudits). The modified operator basis are defined as 
\begin{align*}
    D_{pq}=\chi U_{pq}+\chi^* U^{\dagger}_{pq},
\end{align*}
where $\chi=(1\pm i)/2$ and $U_{pq}$ are usual HW operators. Notice that the modified operators are by construction Hermitian and satisfy the following properties, $D_{00}=\mathbbm{1}$ and ${\rm Tr}[D_{pq}D_{p'q'}]=d \delta_{pp'}\delta_{qq'}$. Therefore, these operators ($\mathbbm{1}$ plus $d^2-1$ operators) form a basis acting on a $d$ dimensional Hilbert space. Thus one can decompose any $d$-dimensional density matrix as
\begin{align*}
    \rho=\frac{1}{d}\sum_{p,q=0}^{d-1}d_{pq} D_{pq},\:\: {\rm with} \quad d_{pq}={\rm Tr} [\rho D_{pq}],
\end{align*}
where the Bloch parameters $d_{pq}$ are real. First, notice that this construction induces a geometry in $\mathbb{R}^{d^2-1}$. 
%Therefore, it does not provide significant advantage over the representation using Gell-Mann operator basis in terms of visualizing Bloch sphere geometry. 
Whereas, our construction induces a geometry in $\mathbb{R}^{(d^2-1)/2}$, which makes it easy to visualize at least in lower dimensions. Also, we find that there is a nontrivial connection between this representation with ours by noticing that $b_{pq}=\chi d_{pq}+\chi^* \omega^{pq} d_{-p,-q}$. Further notice that in our construction, we combine two contributions, $b_{pq}U_{pq}+b_{-p,-q}U_{-p,-q}$, to get $n_j(e^{\mi \theta_j}U_{pq}+e^{-\mi \theta_j}\omega^{pq}U_{-p,-q})$. By plugging one can see that $n_j=d_{pq}+d_{-p,-q}$, whereas solutions for $\theta_j$ comes from 
\begin{align*}
    \chi +\chi^* \omega^{pq}=e^{\mi \theta_j}+e^{-\mi \theta_j}\omega^{pq}.
\end{align*}
%To be mathematically correct, we note that the parameters $(n_j,\theta_j)$ are not exactly coming from above prescription in our construction, rather they are more general. 
It is now easy to see the connection between the present work and the construction presented in Ref. \cite{PhysRevA.94.010301}. Furthermore, the aim of the Ref. \cite{PhysRevA.94.010301} was not to study the geometry induced by their construction, rather they dedicated their study to investigate the witnessing of higher-dimensional entangled states and the discritization of continuous variable systems. Therefore, our study in this perspective can be treated as a companion of the Ref. \cite{PhysRevA.94.010301}.

\section{Conclusion}\label{VIII}
To conclude, we have used the HW operator basis to represent a qutrit state. In doing so, we identified eight independent  parameters consisting of four weight and four angular parameters. We find that the four weight parameters induce a Bloch sphere-like structure in $\mathbb{R}^4$ for qutrits. Further, we have obtained the constraints which must be satisfied for the parametrization to represent a physical qutrit. To understand the geometry of state space, we study its one, two, and three sections in detail. Our study shows that these projections are unlike those studied in the previous literature \cite{Goyal_2016}. 
%This representation seems like a natural extension of the qubit Bloch sphere because of the following properties.
%%\begin{enumerate}
	%\item The one and two-dimensional sections are symmetric with respect to the axes.
	%\item Purity of a state depends on the length of the Bloch vector.
	%\item Rank of the states can be identified to some extent by looking at the distance from the origin of the sphere in $\mathbb{R}^4$.
	%\item The conditions of orthogonality and mutual unbiasedness of two Bloch vectors have a lot of similarities with conditions for the qubit Bloch vectors.
	%\item Hilbert-Schmidt distance between two qutrit states(with same angular parameter values $\theta_i$'s) is proportional to the Euclidean distance in the sphere ($\in \mathbb{R}^4$).
%\end{enumerate}

We have applied our Bloch vector representation to show that there can be a maximum of four MUBs in three dimensions. The characterization of unital maps acting on qutrits is also demonstrated using our representation. We also did a characterization of randomly generated density matrices, when the probability distributions are induced by Hilbert-Schmidt and Bures distances. Lastly, we have mentioned the basic steps required to extend this representation in dimensions greater than three. 

%One of the future works based on extending this work could be to identify the structure of the allowed set of points which represent a physical qutrit state. Another significant work would be to generalize this representation in higher dimensions than three. 

As we have shown in this paper that the geometry of the Bloch sphere limits the existence of the number of MUBs in qubits and qutrits. This approach can be used to study the existence of MUBs in $\mathbb{C}^6$, where the maximum number of MUBs is not known yet \cite{grassl2004sicpovms,PhysRevA.83.062303,Bengtsson2007}. An extension to the characterization of unital maps would be to characterize qutrit entanglement breaking channels similar to qubit entanglement breaking channels \cite{doi:10.1142/S0129055X03001710}. Similar to the characterization of ensembles generated by HS and Bures metric, another interesting study could be to identify the form of the Fubini-Study metric and the corresponding volume element \cite{bengtsson2002cp}. Such an analysis could be useful for sampling pure qutrit states and averaging over them. 

Our sphere representation in $\mathbb{R}^4$ could also have significant applications in studying the dynamics of qudit states and finding the constants of motion in $d$-level systems. It can also be used to detect the entanglement of bipartite systems and identify the reachable states in open system dynamics. We hope that this approach leads to better insight into the study of qudit systems and their dynamics.

{\it Acknowledgement}.-- This work is supported by the Polish National Science Centre through the SONATA BIS project No.2019/34/E/ST2/00369. SS acknowledges funding through PASIFIC program call 2 (Agreement No. PAN.BFB.S.BDN.460.022 with the Polish Academy of Sciences). This project has received funding from the European Union’s Horizon 2020 research and innovation programme under the Marie Sk{\l}odowska- Curie grant agreement No 847639 and from the Ministry of Education and Science. SS also acknowledges the financial support through DEQHOST (APVV-22-0570) and DESCOM (VEGA-2/0183/21) during his stay at IPSAS, Bratislava.

{\em Declaration}.-- a) The authors declare that they have no known competing financial interests or personal relationships that could have appeared to influence the work reported in this paper.\\
b) Data sharing is not applicable to this article as no data sets were generated or analyzed during the current study.

\appendix 
\section{Random Density Matrices in Gell-Mann operator representation}\label{Gell-Mann}
Using the Gell-Mann operator basis also one can write a qutrit state in the following way \cite{Bertlmann_2008}
\begin{align}
	\rho=\frac{1}{3}(\mathbb{I}+\sum_{i=1}^{d^2-1}g_i\Lambda_i),
\end{align}
where $\Lambda_i$ are the Gell-Mann operators in three dimensions and $g_i={\rm Tr}(\Lambda_i\rho)$ form the components of the eight-dimensional(eight-D) Bloch vector $\vec{g}$. The eight Gell-Mann operators in three dimensions contain diagonal, symmetric, and anti-symmetric matrices, but for simplicity, we denote all of them with $\Lambda_i$. 
Using a similar trick as in the case of Weyl operator representation we can get the HS and Bures distribution in terms of the Bloch vector parameters $g_i$ as follows 
\begin{align}
P_{HSG}(\vec{r_g},\alpha_i)=\frac{C_{HSG}}{r^7}F_G(\rho),\:\:P_{BG}(\vec{r_g},\alpha_i)=\frac{C_{BG}F_G(\rho)}{r^7[3-r^2-9\text{Det}(\rho)]\sqrt{\text{Det}(\rho)}},
\end{align}
where $F_G(\rho)=(1/729)(r^2-3)^2(4r^2-3)+[2-2r^2-27\text{Det}(\rho)]\text{Det}(\rho)$ and we have switched to polar representation with $r_g$ being the radial distance in the eight-D Bloch sphere and $\alpha_i$'s being the seven polar angles. $C_{HSG}$ and $C_{BG}$ are constants determined by the normalization. 

As in the Weyl representation, here also the HS distribution is always positive inside the eight-D Bloch sphere irrespective of Det$(\rho)$ being positive or negative. Also, it is non-decreasing with respect to $r_g$. Thus the states chosen are localized at the surface of the Bloch sphere. 

The Bures distribution also behaves similarly to the Weyl representation. It is positive if and only if Det$(\rho)\geq0$ and also it is non-decreasing for Det$(\rho)\geq0$.The singularity in $P_{BG}(\vec{r_g},\alpha_i)$ occurs either at  $[3-r^2-9\text{Det}(\rho)]=0$ or when $\text{Det}(\rho)=0$. The first condition is only possible at or beyond the surface of the eight-D sphere. Whereas, $\text{Det}(\rho)=0$ can happen for rank-1 or rank-2 states, i.e., at the surface of the structure formed by the qutrit states. Thus, $H_{BG}$ is sharply localized at the surface of the convex structure formed by the qutrit states.
\section{Outside of the Ball of radius $r=1/2$}\label{appC}
One can prove that inside the Bloch sphere of radius $r\leq 1/2$, $\Omega$ is positive for all the values of angular parameters $\theta_i$. This can be proven by using the polar coordinate forms of $n_i$'s in Eq.(\ref {a3}), i.e. we replace with $n_1=r \cos\alpha_1, n_2=r \sin\alpha_1\cos\alpha_2,n_3=r \sin\alpha_1\sin\alpha_2\cos\alpha_3,n_4=r \sin\alpha_1\sin\alpha_2\sin\alpha_3\cos\alpha_4$ in Eq.(\ref{a3}), where $\alpha_1,\alpha_2,\alpha_3 $ and $\alpha_4$ are the polar angles. Then $\Omega$ can be written in the following simple form 
\begin{align}\label{a3polar}
	\Omega=1-3r^2+2r^3f(\theta_1,\theta_2,\theta_3,\theta_4,\alpha_1,\alpha_2,\alpha_3,\alpha_4).
\end{align}
where $f \in [-1,1]$ is a function of $\theta_i$'s and $\alpha_i$'s. It is straightforward to see from the above equation that any points inside the Ball of radius, $r=1/2$, corresponds to a physical qutrit. 

Next, we ask whether this boundary is sharp, i.e., if we increase the boundary by $\epsilon<<1$, do all the points on the stretched boundary still corresponds to physical qutrits? If we do little algebra, we find by putting $r=(1/2)+\epsilon$ in the above expression (assuming $\epsilon^2,\epsilon^3 \approx 0$),
\begin{align*}
    \Omega(\epsilon,f)&=1-3\left(\frac{1}{2}+\epsilon\right)^2+2\left(\frac{1}{2}+\epsilon\right)^3f(\bm \theta,\bm \alpha),\\
    &\approx \frac{1}{4}-3\epsilon+\frac{1}{2}\left(\frac{1}{2}+3\epsilon\right)f(\bm \theta,\bm \alpha),\\
    &=\frac{1}{4}\{1+f(\bm \theta,\bm \alpha)\}-3\epsilon\{1-f(\bm \theta,\bm \alpha)\}.
\end{align*}
As $1+f(\bm \theta,\bm \alpha)\geq 0$ always, we look into the second term in the RHS of the last line of the above equation and find that a valid solution ($=0$) exists only when $f(\bm \theta,\bm \alpha)=1$. This means that for arbitrary small $\epsilon$ ($>0$), we no longer have a solid Ball.

%%%%%%%%%%%%%%%%%%%%%%%%%%%%%%%%%%%%%%%%%%%%%%%%%%%%%%%%%%%

%In the qubit Bloch sphere also, the Hilbert-Schmidt distance between two density matrices is proportional to the euclidean distance between them \cite{wilde_2013}.
%\bibliographystyle{ieeetr}
%\bibliographystyle{unsrt}
%\bibliographystyle{elsarticle-harv}
%\bibliographystyle{authordate1}
%\bibliography{blochvectoref}

\begin{thebibliography}{10}

\bibitem{Bertlmann_2008}
R.~A. Bertlmann and P.~Krammer, ``Bloch vectors for qudits,'' {\em Journal of
  Physics A: Mathematical and Theoretical}, vol.~41, p.~235303, may 2008.

\bibitem{RevModPhys.55.855}
U.~Fano, ``Pairs of two-level systems,'' {\em Rev. Mod. Phys.}, vol.~55,
  pp.~855--874, Oct 1983.

\bibitem{Petruccione2012}
E.~Br\"uning, H.~M\"akel\"a, A.~Messina, and F.~Petruccione, ``Parametrizations
  of density matrices,'' {\em Journal of Modern Optics}, vol.~59, pp.~1--20,
  Jan 2012.

\bibitem{KIMURA2003339}
G.~Kimura, ``The bloch vector for n-level systems,'' {\em Physics Letters A},
  vol.~314, no.~5, pp.~339 -- 349, 2003.

\bibitem{kimura2004blochvector}
G.~Kimura and A.~Kossakowski, ``The bloch-vector space for n-level systems --
  the spherical-coordinate point of view,'' 2004.

\bibitem{2006quant.ph..2065K}
S.~{Kryszewski} and M.~{Zachcial}, ``{Alternative representation of N $\times$
  N density matrix},'' {\em arXiv e-prints}, pp.~quant--ph/0602065, Feb. 2006.

\bibitem{Menda__2006}
I.~P. Menda{\v{s}}, ``The classification of three-parameter density matrices
  for a qutrit,'' {\em Journal of Physics A: Mathematical and General},
  vol.~39, pp.~11313--11324, aug 2006.

\bibitem{Goyal_2016}
S.~K. Goyal, B.~N. Simon, R.~Singh, and S.~Simon, ``Geometry of the generalized
  bloch sphere for qutrits,'' {\em Journal of Physics A: Mathematical and
  Theoretical}, vol.~49, p.~165203, mar 2016.

\bibitem{Eltschka_2021}
C.~Eltschka, M.~Huber, S.~Morelli, and J.~Siewert, ``The shape of
  higher-dimensional state space: Bloch-ball analog for a qutrit,'' {\em
  Quantum}, vol.~5, p.~485, jun 2021.

\bibitem{10.5555/2011406.2011407}
P.~Kurzy\'{n}ski, ``Multi-bloch vector representation of the qutrit,'' {\em
  Quantum Info. Comput.}, vol.~11, p.~361–373, May 2011.

\bibitem{PhysRevA.93.062126}
P.~Kurzy\ifmmode~\acute{n}\else \'{n}\fi{}ski, A.~Ko\l{}odziejski,
  W.~Laskowski, and M.~Markiewicz, ``Three-dimensional visualization of a
  qutrit,'' {\em Phys. Rev. A}, vol.~93, p.~062126, Jun 2016.

\bibitem{Vourdas_2004}
A.~Vourdas, ``Quantum systems with finite hilbert space,'' {\em Reports on
  Progress in Physics}, vol.~67, pp.~267--320, feb 2004.

\bibitem{PhysRevA.94.010301}
A.~Asadian, P.~Erker, M.~Huber, and C.~Kl\"ockl, ``Heisenberg-weyl observables:
  Bloch vectors in phase space,'' {\em Phys. Rev. A}, vol.~94, p.~010301, Jul
  2016.

\bibitem{Bandyopadhyay2002}
{S. Bandyopadhyay}, {P.O. Boykin}, {V. Roychowdhury}, and {F. Vatan}, ``A new
  proof for the existence of mutually unbiased bases,'' {\em Algorithmica},
  vol.~34, pp.~512--528, Nov 2002.

\bibitem{1997PhDT.......232G}
D.~{Gottesman}, {\em {Stabilizer codes and quantum error correction}}.
\newblock PhD thesis, California Institute of Technology, Jan. 1997.

\bibitem{PhysRevA.66.032319}
R.~A. Bertlmann, H.~Narnhofer, and W.~Thirring, ``Geometric picture of
  entanglement and bell inequalities,'' {\em Phys. Rev. A}, vol.~66, p.~032319,
  Sep 2002.

\bibitem{chang2018separability}
J.~Chang, M.~Cui, T.~Zhang, and S.-M. Fei, ``Separability criteria based on
  heisenberg--weyl representation of density matrices,'' {\em Chinese Physics
  B}, vol.~27, no.~3, p.~030302, 2018.

\bibitem{Cotfas_2012}
N.~Cotfas and D.~Dragoman, ``Properties of finite gaussians and the
  discrete-continuous transition,'' {\em Journal of Physics A: Mathematical and
  Theoretical}, vol.~45, p.~425305, oct 2012.

\bibitem{PhysRevA.74.032327}
B.~Baumgartner, B.~C. Hiesmayr, and H.~Narnhofer, ``State space for two qutrits
  has a phase space structure in its core,'' {\em Phys. Rev. A}, vol.~74,
  p.~032327, Sep 2006.

\bibitem{PhysRevLett.114.250403}
A.~Asadian, C.~Budroni, F.~E.~S. Steinhoff, P.~Rabl, and O.~G\"uhne,
  ``Contextuality in phase space,'' {\em Phys. Rev. Lett.}, vol.~114,
  p.~250403, Jun 2015.

\bibitem{PhysRevLett.70.1895}
C.~H. Bennett, G.~Brassard, C.~Cr\'epeau, R.~Jozsa, A.~Peres, and W.~K.
  Wootters, ``Teleporting an unknown quantum state via dual classical and
  einstein-podolsky-rosen channels,'' {\em Phys. Rev. Lett.}, vol.~70,
  pp.~1895--1899, Mar 1993.

\bibitem{doi:10.1142/S0129055X03001710}
M.~B. Ruskai, ``Qubit entanglement breaking channels,'' {\em Reviews in
  Mathematical Physics}, vol.~15, no.~06, pp.~643--662, 2003.

\bibitem{PhysRevA.70.062101}
K.~S. Gibbons, M.~J. Hoffman, and W.~K. Wootters, ``Discrete phase space based
  on finite fields,'' {\em Phys. Rev. A}, vol.~70, p.~062101, Dec 2004.

\bibitem{wilde_2013}
M.~M. Wilde, {\em Quantum Information Theory}.
\newblock Cambridge University Press, 2013.

\bibitem{PhysRevA.17.1249}
J.~E. Harriman, ``Geometry of density matrices. {I.} definitions, {$N$}
  matrices and $1$ matrices,'' {\em Phys. Rev. A}, vol.~17, pp.~1249--1256, Apr
  1978.

\bibitem{sharma2019fine}
G.~Sharma, S.~Sazim, and S.~Mal, ``Role of fine-grained uncertainty in
  determining the limit of preparation contextuality,'' {\em Phys. Rev. A},
  vol.~104, p.~032424, Sep 2021.

\bibitem{904522}
C.~{King} and M.~B. {Ruskai}, ``Minimal entropy of states emerging from noisy
  quantum channels,'' {\em IEEE Transactions on Information Theory}, vol.~47,
  pp.~192--209, Jan 2001.

\bibitem{BETHRUSKAI2002159}
M.~{Beth Ruskai}, S.~Szarek, and E.~Werner, ``An analysis of
  completely-positive trace-preserving maps on m2,'' {\em Linear Algebra and
  its Applications}, vol.~347, no.~1, pp.~159 -- 187, 2002.

\bibitem{zyczkowski2011generating}
K.~{\.Z}yczkowski, K.~A. Penson, I.~Nechita, and B.~Collins, ``Generating
  random density matrices,'' {\em Journal of Mathematical Physics}, vol.~52,
  no.~6, p.~062201, 2011.

\bibitem{Zyczkowski_2001}
K.~{\.Z}yczkowski and H.-J. Sommers, ``Induced measures in the space of mixed
  quantum states,'' {\em Journal of Physics A: Mathematical and General},
  vol.~34, pp.~7111--7125, aug 2001.

\bibitem{HALL1998123}
M.~J. Hall, ``Random quantum correlations and density operator distributions,''
  {\em Physics Letters A}, vol.~242, no.~3, pp.~123--129, 1998.

\bibitem{bures1969extension}
D.~Bures, ``An extension of kakutani's theorem on infinite product measures to
  the tensor product of semifinite w*-algebras,'' {\em Transactions of the
  American Mathematical Society}, vol.~135, pp.~199--212, 1969.

\bibitem{uhlmann1992metric}
A.~Uhlmann, ``The metric of bures and the geometric phase,'' in {\em Groups and
  related Topics}, pp.~267--274, Springer, 1992.

\bibitem{hubner1992explicit}
M.~H{\"u}bner, ``Explicit computation of the bures distance for density
  matrices,'' {\em Physics Letters A}, vol.~163, no.~4, pp.~239--242, 1992.

\bibitem{grassl2004sicpovms}
M.~{Grassl}, ``{On SIC-POVMs and MUBs in Dimension 6},'' {\em arXiv e-prints},
  pp.~quant--ph/0406175, June 2004.

\bibitem{Goyeneche_2013}
D.~Goyeneche, ``Mutually unbiased triplets from non-affine families of complex
  hadamard matrices in dimension 6,'' {\em Journal of Physics A: Mathematical
  and Theoretical}, vol.~46, p.~105301, feb 2013.

\bibitem{PhysRevA.83.062303}
P.~Raynal, X.~L\"u, and B.-G. Englert, ``Mutually unbiased bases in six
  dimensions: The four most distant bases,'' {\em Phys. Rev. A}, vol.~83,
  p.~062303, Jun 2011.

\bibitem{Bengtsson2007}
I.~Bengtsson, W.~Bruzda, {\AA}.~Ericsson, J.-{\AA}. Larsson, W.~Tadej, and
  K.~{\.Z}yczkowski, ``Mutually unbiased bases and hadamard matrices of order
  six,'' {\em Journal of mathematical physics}, vol.~48, no.~5, p.~052106,
  2007.

\bibitem{bengtsson2002cp}
I.~Bengtsson, J.~Br{\"a}nnlund, and K.~{\.Z}yczkowski, ``Cp$^n$, or,
  entanglement illustrated,'' {\em International Journal of Modern Physics A},
  vol.~17, no.~31, pp.~4675--4695, 2002.

\end{thebibliography}

\end{document}